\documentclass[sigplan]{acmart}
\renewcommand\footnotetextcopyrightpermission[1]{} 
\usepackage{makecell}
\usepackage{balance}
\usepackage{algorithmicx}
\usepackage{amsmath}
\theoremstyle{definition}
\newtheorem{theorem}{Theorem}
\usepackage{graphicx}
\usepackage{url}
\usepackage{pifont}
\usepackage{subcaption}
\usepackage[linesnumbered,vlined,ruled]{algorithm2e}
\usepackage{multirow}
\graphicspath{{./figures/}}

\newtheorem{innercustomthm}{Theorem}
\newenvironment{customthm}[1]
  {\renewcommand\theinnercustomthm{#1}\innercustomthm}
  {\endinnercustomthm}
  
\newtheorem{innerlemma}{Lemma}
\newenvironment{customlemma}[1]
  {\renewcommand\theinnerlemma{#1}\innerlemma}
  {\endinnerlemma}

\newcommand{\systemname}{GraphZero}

\title{\systemname: Breaking Symmetry for Efficient Graph Mining}

\begin{document}

\author{Daniel Mawhirter$^*$,\hspace{1em} Sam Reinehr$^*$,\hspace{1em} Connor Holmes$^*$,\hspace{1em} Tongping Liu$^{\mathsection}$,\hspace{1em} Bo Wu$^*$ }
\affiliation{%
 \institution{Colorado School of Mines$^*$ \hspace{0.3em} University of Massachusetts at Amherst$^\mathsection$}
}
\affiliation{%
  \institution{\{dmawhirt,\hspace{0.2em}swreinehr,\hspace{0.2em}cholmesd\}@mymail.mines.edu} tongping@umass.edu, bwu@mines.edu
}

\begin{abstract}
Graph mining for structural patterns is a fundamental task in many applications. Compilation-based graph mining systems, represented by AutoMine, generate specialized algorithms for the provided patterns and substantially outperform other systems. However, the generated code causes substantial computation redundancy and the compilation process incurs too much overhead to be used online, both due to the inherent symmetry in the structural patterns. 

In this paper, we propose an optimizing compiler, GraphZero, to completely address these limitations through symmetry breaking based on group theory. GraphZero implements three novel techniques. First, its schedule explorer efficiently prunes the schedule space without missing any high-performance schedule. Second, it automatically generates and enforces a set of restrictions to eliminate computation redundancy. Third, it generalizes orientation, a surprisingly effective optimization that was mainly used for clique patterns, to apply to arbitrary patterns. Evaluated on multiple graph mining applications and complex patterns with 7 real-world graph datasets, GraphZero demonstrates up to 40X performance improvement and up to 197X reduction on schedule generation overhead over AutoMine.
\end{abstract}

\maketitle

\section{Introduction} \label{sec:intro}

Graph data is ubiquitous and used in numerous domains thanks to its flexibility. Graph mining searches for structural patterns in large-scale graphs, which is important for Bioinformatics~\cite{Hocevar:Bioinfo2006,Wernicke:Bioinfo2006}, social networks~\cite{Newman03thestructure,Yang:ASONAM18}, fraud detection~\cite{Akoglu:CoRR2014,sjtree}, and so on. 
Triangle counting represents a simple graph mining task, where the pattern is a triangle and the goal is to count all its instances in an input graph. While triangle counting is much more costly to solve than many graph traversal problems, mining larger patterns (e.g., size-5 cliques) involves even more complex algorithms and may need hours or days to finish.


Graph mining has attracted significant attention in the data analytics community. A common approach is to design efficient algorithms for individual patterns~\cite{Chu:TKDD2012,Zhu:ICDM18,Ahmed:ICDM2015}. IEEE, Amazon and MIT organize an annual challenge to rank the submitted implementations for triangle counting~\cite{graphchallenge}. However, this approach is not scalable for general graph mining as there exist too many patterns of dramatically varying structures. Moreover, the number of patterns increases exponentially in the pattern size. For example, there exist 112 size-6 patterns but 853 size-7 patterns. 

Many software systems have been proposed to mine arbitrary patterns~\cite{grami,scalemine,distgraph,gspan,arabesque,gminer,Kaleido,rstream}. Despite their generality, these systems usually have poor performance due to two reasons. First, they implement a generic algorithm that handles arbitrary patterns but does not perform particularly well for any of them. It is why manual implementations of specialized algorithms for specific patterns are still popular. Second, they need to run some kind of isomorphism testing online to verify whether a produced embedding matches the pattern of interest, which may incur substantial overhead for non-trivial patterns.


The compilation approach for graph mining has unique advantages as shown by systems like EmptyHeaded~\cite{aberger2017emptyheaded}, GraphFlow~\cite{kankanamge2017graphflow} and AutoMine~\cite{automine}. They generate specialized algorithms for the given pattern and compile them to efficient low-level code. The code has a nested-loop structure, each growing the embedding by one vertex towards the pattern through set operations on the neighbor lists (e.g, set intersection). If an embedding is identified by the innermost loop, it guarantees to match the pattern and avoids online isomorphism testing. 


AutoMine represents the state of the art for compilation-based graph mining systems and often substantially outperforms other systems. But it has three problems. The first problem is that its generated code incurs significant computation redundancy because it may identify the same instance multiple times.
For a size-7 chordal cycle, which is a clique with one absent edge, AutoMine identifies each of its embeddings 120 times. The second problem is AutoMine's slow compilation speed. Provided with a pattern, AutoMine uses a brute-force approach to enumerate all possible schedules. As a result, its search algorithm may unnecessarily traverse schedules that generate the same code. If used as an offline compiler, the compilation overhead may be acceptable especially because most real-world applications only mine small patterns. But it may be problematic if AutoMine is used as a just-in-time compiler for dynamic queries where the overhead lies on the critical path. Finally, AutoMine, as well as other graph mining systems, can only apply the orientation optimization~\cite{tricore,tri} to clique patterns, which only allows edge traversals from lower-degree vertices to higher-degree vertices. However, it is one of the most effective optimizations for graph mining. Triangle counting, for instance, enjoys up to tens of times performance improvements from this technique alone~\cite{Shun:ICDE15}.

Addressing the three problems faces substantial challenges. First, patterns and their inner structures have symmetry. 
Consider the rectangle pattern.
All four vertices are initially equivalent due to the four-way rotational symmetry.
After fixing one vertex, there is still a two-way mirrored symmetry between the two vertices attached to the first, leading to a total multiplicity of 8.
The symmetry leads to different ways to map the pattern to the vertices in the graph. However, the set operations can only recognize the topology and are hence not sufficient to break symmetry. Second, the schedule space can be enormous, and different schedules may produce programs that differ dramatically in performance. Although we can naively prune the schedule space by setting an upper bound on the number of explored schedules, this may miss highly performant schedules. Third, the orientation optimization can be applied to clique patterns because they are perfectly symmetric, meaning that any pair of vertices can be exchanged yet still preserving the pattern. Most patterns do not have this property, so it is unclear whether they can also benefit from this optimization.

In this paper, we present GraphZero, an optimizing compiler for graph mining that systematically addresses these challenges based on symmetry breaking using group theory. GraphZero contains three key components: a schedule explorer, a redundancy optimizer, and an orientation optimizer. Given the patterns of interest, the schedule explorer can search for an optimized schedule in a few milliseconds. The redundancy optimizer automatically generates a set of restrictions for the found schedule and enforces the restrictions on it to completely eliminate computation redundancy. The orientation optimizer successfully generalizes the application of the orientation optimization to arbitrary patterns.

The fundamental idea for GraphZero to break symmetry is to use more than the topology information, such as vertex IDs. We observe that the different ways to identify the same embedding are essentially automorphisms, which are isomorphisms from the instance to itself. The number of automorphisms determines the degree of redundancy. By enforcing a set of automatically generated restrictions between discovered vertices, GraphZero
reduces the number of automorphisms to one. Consequently, the generated code
identifies each instance exactly once. GraphZero extends this idea to schedule pruning and
only discovers one of the schedules that correspond to the same automorphism group. It does not miss the optimal schedule because all these schedules
order the set operations in the same way and hence have the same performance. Finally, GraphZero enjoys a byproduct of the enforced restrictions, which is that for each restriction it can also restrict an edge traversal order and thus generalizes the orientation optimization.

We have extensively evaluated all the three components of GraphZero by comparing it with AutoMine on 7 real-world graphs of different scales.
Our experimental results show that GraphZero achieves up to 8X, 10X, 22X, and 40X performance improvements for Motif-4 counting, Motif-5 counting, Pentagon mining and size-7 chordal cycle mining.
We observe that the more complex the pattern is, the more benefit GraphZero provides.
GraphZero generates high-performance schedules up to 197X faster for multiple complex patterns.
Moreover, GraphZero's generalized orientation optimization produces up to 40x performance improvement over AutoMine on non-clique patterns.



We make the following contributions in this paper: 1) We reveal that the inherent symmetry in graph patterns is the fundamental reason for both AutoMine's computation redundancy and its slow compilation speed; 2) We propose to use group theory to break symmetry through automatically generated and enforced restrictions, which not only completely eliminates redundancy in the generated code but also substantially prunes the schedule search space; 3) We generalize the orientation optimization to arbitrary patterns by leveraging the generated restrictions; 4) We present an optimizing compiler that integrates the proposed techniques to substantially outperform AutoMine with a much faster compilation speed.

\section{Background and Motivation} \label{sec:moti}


\subsection{AutoMine basics}

\begin{figure*}[ht] \centering
\begin{subfigure}[t]{0.65\textwidth} \centering
\includegraphics[height=0.85in]{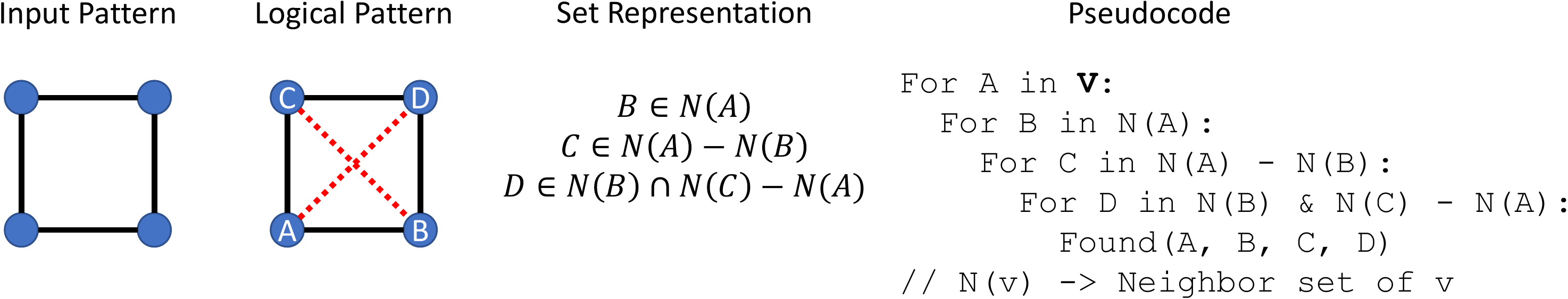}
\caption{Scheduling Phase} 
\end{subfigure} ~
\begin{subfigure}[t]{0.30\textwidth} \centering
\includegraphics[height=0.85in]{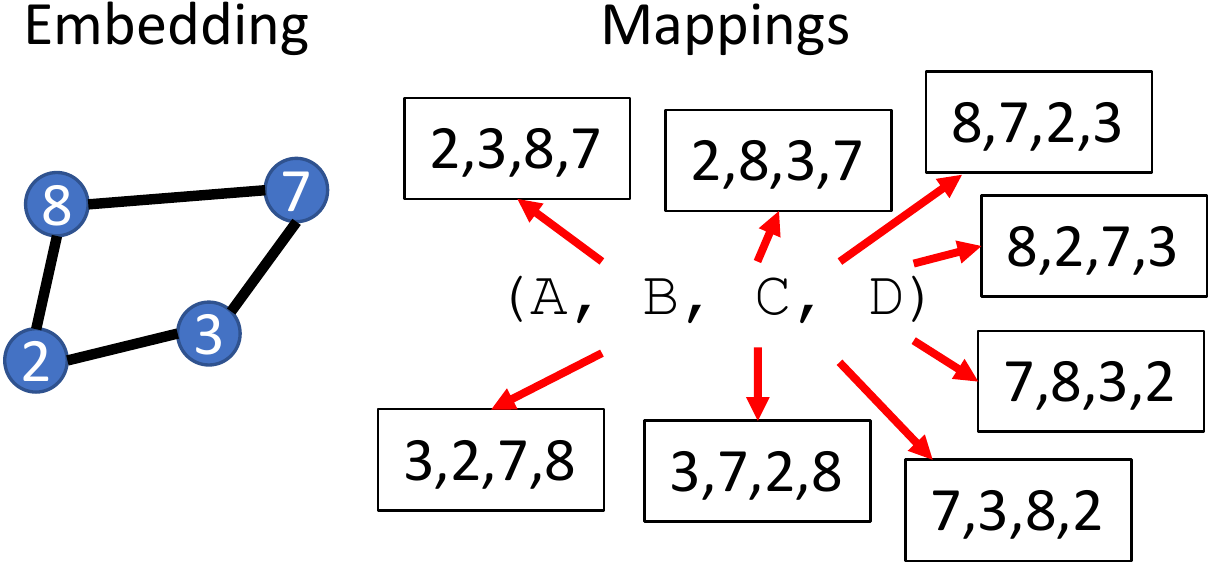}
\caption{Mining Phase} 
\end{subfigure}
\caption{The full process of AutoMine.}
\label{fig:automine-process} \end{figure*}



The AutoMine~\cite{automine} system introduces a \emph{topological} compiler for graph mining. Unlike the general graph mining systems that preceded it~\cite{arabesque,rstream}, AutoMine uses an offline compilation process to analyze the given pattern and generate code to identify all its instances in an input graph. Interestingly, the paper shows that any arbitrary connected pattern can be identified by a series of \emph{set intersection} and \emph{set subtraction} operations called a {\em schedule}. This set-centric representation turns out to be powerful and substantially reduces memory consumption compared with other systems, leading to hundreds of times performance improvements over other systems for many patterns and real-world graphs.

Figure~\ref{fig:automine-process} (a) demonstrates the basic workflow of AutoMine. Given a pattern of interest (i.e., a rectangle in the example), it assigns a distinct label to each of the vertices in the pattern represented by $\{A, B, C, D\}$. AutoMine then generates a colored complete graph (i.e., the logical pattern) with existing edges in black and absent edges in red. AutoMine then produces a discovery sequence of the labeled vertices, making sure that any vertex in the sequence except the first one is connected to at least one earlier vertex through a black edge. A schedule can then be naturally generated following this sequence with black edges encoded by set intersections and red edges encoded by set subtractions. For example, $D$ is connected to both $B$ and $C$ through black edges and $A$ through a red edge, AutoMine understands that $D$ should be in the intersection of $B$'s and $C$'s neighbor sets but not in $A$'s neighbor set. Finally, AutoMine compiles the schedule into a nested loop structure, the inner-most loop of which identifies the instances of the pattern. Since the discovery sequence determines a schedule, we use a total order on the label set to represent the schedule. For example, the schedule shown in Figure~\ref{fig:automine-process} (a) can be represented by $(A, B, C, D)$.

\subsection{Problems and Challenges}

\subsubsection{Computation redundancy}
Every graph mining system needs a way to test isomorphism, that is, determine what pattern a particular subgraph matches. AutoMine embeds the isomorphism test in the generated code itself, using the nested loop structure to filter embeddings down to those that match the desired structural pattern. Doing so discovers the vertices in a particular order, what we call a \emph{mapping} from the pattern's labeled vertices to the embedded vertices in the graph. However, for the same embedding AutoMine may map the label set to its vertices in different ways. Figure~\ref{fig:automine-process} (b) shows that there exists eight different mappings for a rectangle instance, leading to eight times over-counting (also called multiplicity) and computation redundancy. 

\begin{figure} \centering
\includegraphics[width=.45\textwidth]{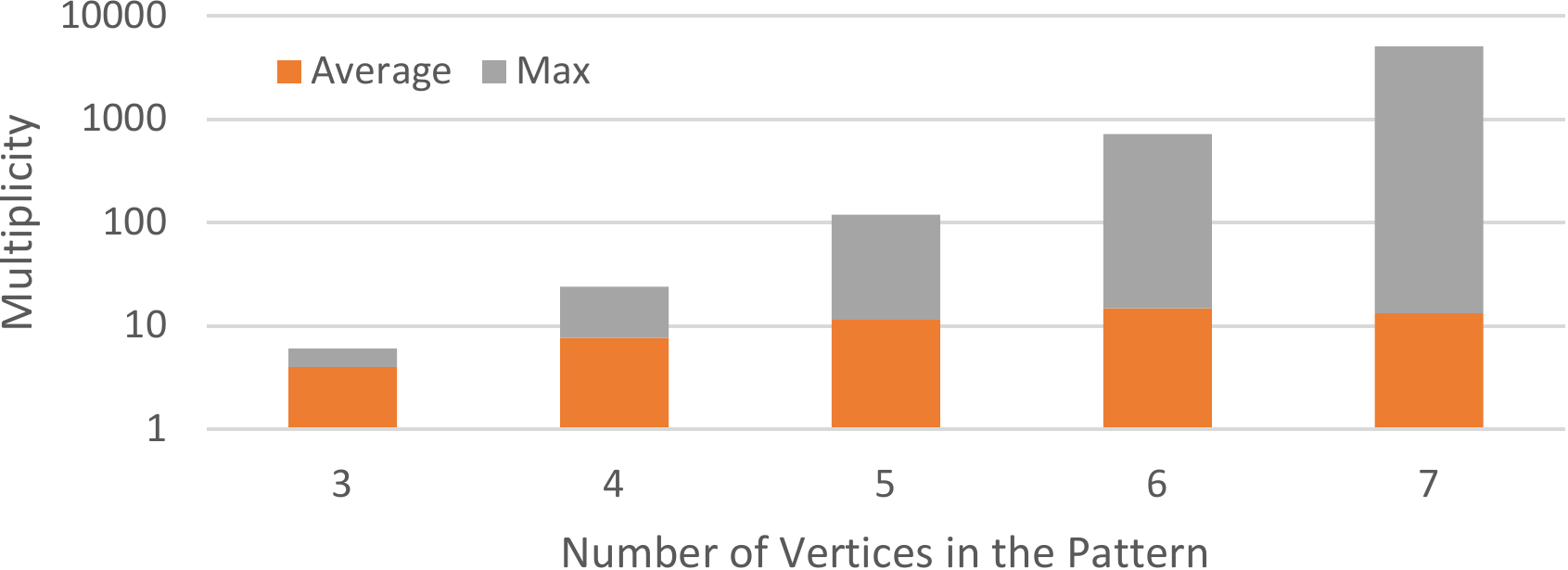}
\caption{The computation redundancy explosion problem}
\label{fig:multiplicity} \end{figure}

AutoMine partially solves the problem by introducing the concept of root symmetry. It checks whether a schedule's first two vertices are equivalent and if so only processes that first "root" edge in one direction. For many patterns, this successfully cuts the computation redundancy in half. However, as Figure~\ref{fig:multiplicity} shows, the number of possible mappings explodes with the number of vertices in the pattern. Ideally, the system should only identify each instance exactly once and so completely eliminate computation redundancy.

\paragraph{Challenges}
The scheduling approach of AutoMine is fundamentally unequipped to handle this multiplicity problem. All the mappings correspond to the same topology and cannot be differentiated by only set operations. As such, the symmetries in a pattern induce the same multiplicity in any possible schedule for that pattern. The symmetry may express itself in different ways in different schedules, as the vertices that can satisfy a particular mapping depend on both the schedule and the previously explored vertices. This diversity is a key reason the problem is both challenging to handle and valuable to solve. Our approach must find all the symmetries and a way to break them to make the mapping unique.

\subsubsection{Brute-force Schedule Generation}
AutoMine uses a naive approach to schedule generation.
It first enumerates all the possible schedules and removes infeasible ones (i.e., there exists a vertex in the discovery sequence which is not connected to any previous vertex through a black edge).
A performance model then evaluates all of the feasible schedules and selects a final one.
This process is inefficient because many schedules generate the same code and thus have the same performance.
AutoMine takes over 147 seconds to generate schedules for 7-vertex motif counting.
When used as a just-in-time compiler, this latency can be problematic because the schedule search overhead lies on the critical path.






\paragraph{Challenges}
The compilation process introduces a chicken-and-egg problem: we want to avoid an exhaustive search of the scheduling space while still exploring all the relevant points, but we do not know the relevant points until after exploring the space. We can run isomorphism testing to filter out schedules that produce the same code, but the testing itself involves an exponential algorithm. Moreover, estimating the performance of schedules is more complex if we want to reduce the computation redundancy.

\subsubsection{Generalization of Orientation Optimization}

Orientation is a popular optimization for triangle counting~\cite{Shun:ICDE15,tricore,tri}. By only allowing higher-degree vertices to be discovered after lower-degree vertices to form triangles, the optimization effectively prunes half of the edges (i.e., the ones from higher-degree vertices to lower-degree vertices) but still preserves the structure of all triangle embeddings. But it can achieve significant speedups much larger than 2X because the time complexity of triangle counting grows super-linearly with the maximum degree. AutoMine observes that the clique pattern is perfectly symmetric and can hence enforce an order to discover the vertices in each embedding from low to high degree. It can then apply the orientation optimization to any-size clique pattern. However, most patterns are not perfectly symmetric and cannot enjoy the benefit from the orientation optimization in AutoMine.

\paragraph{Challenges}

Generalizing the orientation optimization to cover arbitrary patterns is difficult. Enforcing an order for vertex discovery based on degree conceptually prunes edges from the graph data. Doing so may miss embeddings of interest because it may be necessary to discover a vertex from a higher-degree vertex but the edge is not present. While manual application of the optimization for specific patterns is possible, automating it in a compiler entails a systematic approach.

\section{Overview of \systemname}

\begin{figure*}
    \centering
    \includegraphics[width=\textwidth]{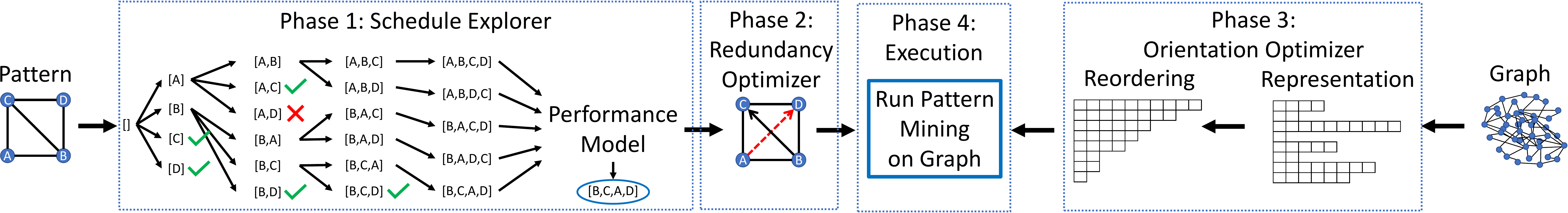}
    \caption{Overview of GraphZero}
    \label{fig:overview}
\end{figure*}

The overall workflow of GraphZero has four stages as shown in Figure~\ref{fig:overview}. Taking a pattern as the input, the schedule explorer searches only part of the schedule space while guaranteeing that the optimal schedule is discovered. Based on a performance model, the schedule explorer estimates the performance of the searched schedules and produces a high-performance schedule for the redundancy optimizer. The redundancy optimizer then generates a set of restrictions and enforces them on the schedule to perform redundancy-free computation. The orientation optimizer, which is optional, reindexes the vertices in the input graph based on decreasing degree and leverages the generated restrictions to conceptually prune the corresponding edges from higher-degree to lower-degree vertices while keeping the entire graph. Finally, the optimized schedule can run on the reindexed graph to mine the input pattern. Although the discussion focuses on one pattern for simplicity, if provided multiple patterns GraphZero merges the schedules to only generate one mining program.


\section{Redundancy-Free Code Generation} \label{sec:reduction-intuition}

We delay the discussion of schedule generation to the next section because the insights from this section will be needed to accelerate that process. 
In this section, we first explain the essential reason for computation redundancy and build the connection between redundancy and automorphisms, which is the key concept from group theory to address the problem. 
We next describe the approach to completely eliminating computation redundancy through automatically generated and enforced restrictions, followed by a method to minimize the overhead to implement the restrictions in the generated schedule. 
Finally, we show what the output looks like with and without these restrictions.

\subsection{Computation Redundancy and Automorphism}

The computation redundancy problem caused by over-counting roots from the symmetry in the pattern. 
For example, given an embedding of the rectangle pattern, each of the four vertices in the embedding can be indistinguishably mapped to the first label in the schedule. 
Once the first vertex in the embedding is mapped, both its neighbor vertices can be mapped to the second label of the schedule. 
The multiplicative mapping choices result in the eight times over-counting. 
Note that selecting a better schedule does not solve the problem because the symmetry cannot be broken by only set operations.

A mapping represents a one-to-one relationship between the labels in the pattern and the vertices in the embedding. 
If we give a total ordering to the embedding vertices, each mapping can be represented by a total ordering of the labels.
Consider the example in Figure~\ref{fig:automine-process}. 
We have $(A,B,C,D)$ to denote the schedule, so a total ordering of the embedding vertices can be $(2,3,8,7)$. 
The mapping from $(A, B, C, D)$ to $(2, 8, 3, 7)$ can then be represented by the total ordering of the labels: $(A, C, B, D)$.
We call a total ordering of the labels that corresponds to a mapping a valid ordering.
Not all total orderings correspond to a mapping.
For instance, $(A, C, D, B)$ is not a valid ordering because $C$ and $D$ are connected in the pattern, but the embedding vertices 3 and 8 are not.

Each valid ordering is essentially a permutation of the total ordering representing the schedule~\footnote{A total ordering is a set plus the relation on the set, but we treat it as a label array to simplify the discussion}.
Each such permutation function is called an automorphism in group theory, which is an isomorphism from the pattern to itself and hence preserves the structure of the pattern. 
If an automorphism repositions a label $A$ to the original position of another label $B$ (e.g., $(A, B, C, D) \rightarrow (B, A, C, D)$), we say that the automorphism moves $A$ to $B$ for simplicity.
All the automorphisms for the same pattern form an automorphism group.
The size of the automorphism group determines the number of times for over-counting.
Ideally, we want to reduce the size of the group to one to completely eliminate computation redundancy.

\subsection{Breaking Symmetry through Restrictions}

The key idea is that we can enforce restrictions on the IDs of the embedding vertices to break the symmetry inherent in the pattern.
Vertices are not reused within the same embedding, so there is a natural total ordering on the IDs.
Consider the rectangle pattern in Figure~\ref{fig:automine-process} again.
We have observed that the first label in the schedule can be mapped to any of the four embedding vertices.
The restriction we enforce is that the first mapped vertex should have the highest ID among the four embedding vertices. 
Doing this effectively reduces the over-counting by four times because only two mappings, namely $(A, B, C, D) \rightarrow (8, 7, 2, 3)$ and $(A, B, C, D) \rightarrow (8, 2, 7, 3)$ respect the restrictions.
We then enforce another restriction that the ID of the second mapped vertex should be larger than that of the third mapped vertex.
The end result is that the only way to identify the embedding is through the mapping $(A, B, C, D) \rightarrow (8, 7, 2, 3)$, which completely solves the over-counting problem.

We make two observations on the example.
First, the restrictions are each a transitive binary relation (i.e., larger than) between the IDs of two embedding vertices.
Second, we can use a binary ordering relation on the labeled vertices of the pattern to represent the restriction on the embedding vertices.
As such, the partial ordering $\{(A, B), (A, C), (A, D), (B, C)\}$ should be sufficient to generate all restrictions. 

Algorithm~\ref{alg:multiplicity_reduce} generalizes the idea and generates such a partial ordering for an arbitrary pattern with a given schedule.
The algorithm at the beginning computes the automorphism group by trying all permutations of the vertex labels and filtering out invalid ones, which do not preserve the pattern. 
It then iterates through all labeled vertices in the schedule.
In each iteration, it determines all the labeled vertices indistinguishable from the traversed vertex $v$ after prior partial ordering. For each of of these labeled vertices $x(v)$, which the automorphism $x$ maps $v$ to, the algorithm adds a binary relation $(v, x(v))$ to the resultant partial ordering. At the end of the for loop, only automorphisms that do not move the traversed vertices are used to generate binary relations for the next traversed vertex.  




\begin{algorithm}
\SetKwInOut{Input}{input}
\SetKwInOut{Output}{output}
\SetKwProg{procedure}{Procedure}{}{}
\DontPrintSemicolon
\Input{$ $ $P:$ the pattern.}
\Input{$ $ $S:$ the schedule.}
\Output{$ $ $L:$ a partial ordering on the labels}
\Begin{
    $Aut \gets$ all the automorphisms of $P$\\
    $L\gets$ an empty partial ordering.\\
    \tcp{iterate through these in order, $S[0], S[1],\cdots$}
    \For{$v$ in  $S$}{
        \tcp{stabilized\_aut contains all automorphisms that do not move the vertices we have iterated over}
        $stabilized\_ aut\gets []$\\
        \For{$x \in Aut$}{
            \If{$x(v)= v$}{
            \tcp{x does not move v}
                add $x$ to $stabilized\_aut$}
            \Else{
                \tcp{$x$ moves v and indicates a binary ordering relation}
                add $(v, x(v))$ to L, if not present}}
            $Aut\gets stabilized\_aut$}}
\caption{Restriction generation algorithm}
\label{alg:multiplicity_reduce}
\end{algorithm}


The following theorem allows us to use these relations to uniquely discover each embedding exactly once.

\begin{theorem}
\label{theorem:multiplicity_reduce}: 
For an arbitrary pattern $P$ and its schedule $S$, let $L$ be the partial ordering on the label set in $P$ generated by Algorithm~\ref{alg:multiplicity_reduce}.
Given an instance of $P$, denoted by $e$, in a graph $G$ there exists exactly one mapping $M$ from $P$ to $e$ if for each binary ordering relation $(S[i], S[j])$ in $L$, we add a restriction $M(S[i]).id > M(S[j]).id$. 

\end{theorem}
We omit the proof here but include it in the appendix. The proof has two steps. We first show that there is at most one mapping that follows all the restrictions, as otherwise there is a restriction that is not enforced. We then show there is at least one mapping that satisfies the restrictions, if one exists at all.  

Algorithm \ref{alg:multiplicity_reduce} and Theorem \ref{theorem:multiplicity_reduce} demonstrate the possibility to completely remove computation redundancy. Implementing the restrictions in the code generator is trivial. The compiler needs to insert bound checks when generating a for loop as demonstrated in Figure~\ref{fig:alg-rect} (a).

\subsection{Minimizing the Overhead of Enforcing Restrictions}

\begin{figure*}[ht] \centering
    \begin{subfigure}[t]{0.45\textwidth} \centering
        \includegraphics[height=1.8in]{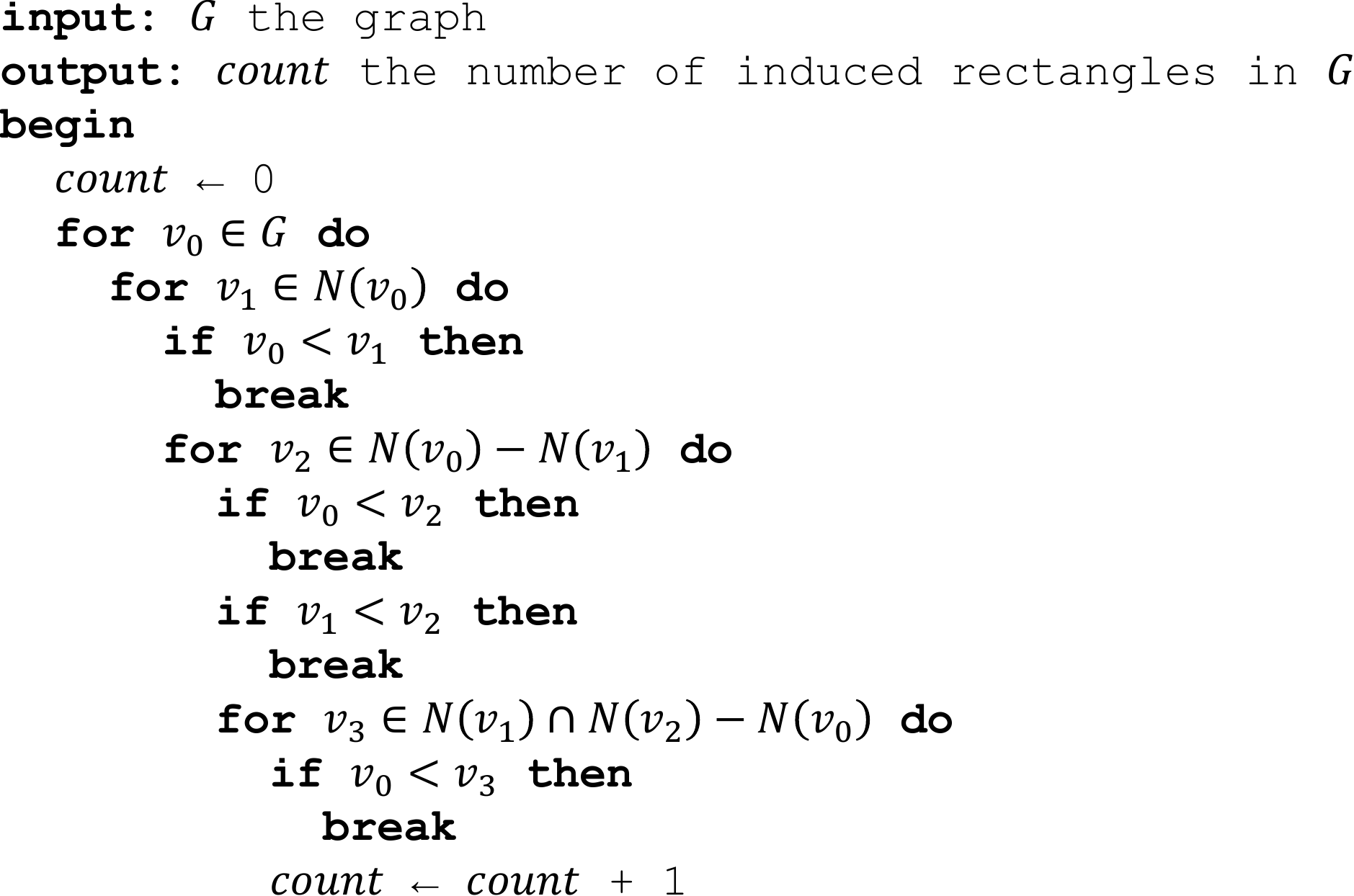}
        \caption{Applying All Restrictions} 
    \end{subfigure} ~
    \begin{subfigure}[t]{0.45\textwidth} \centering
        \includegraphics[height=1.8in]{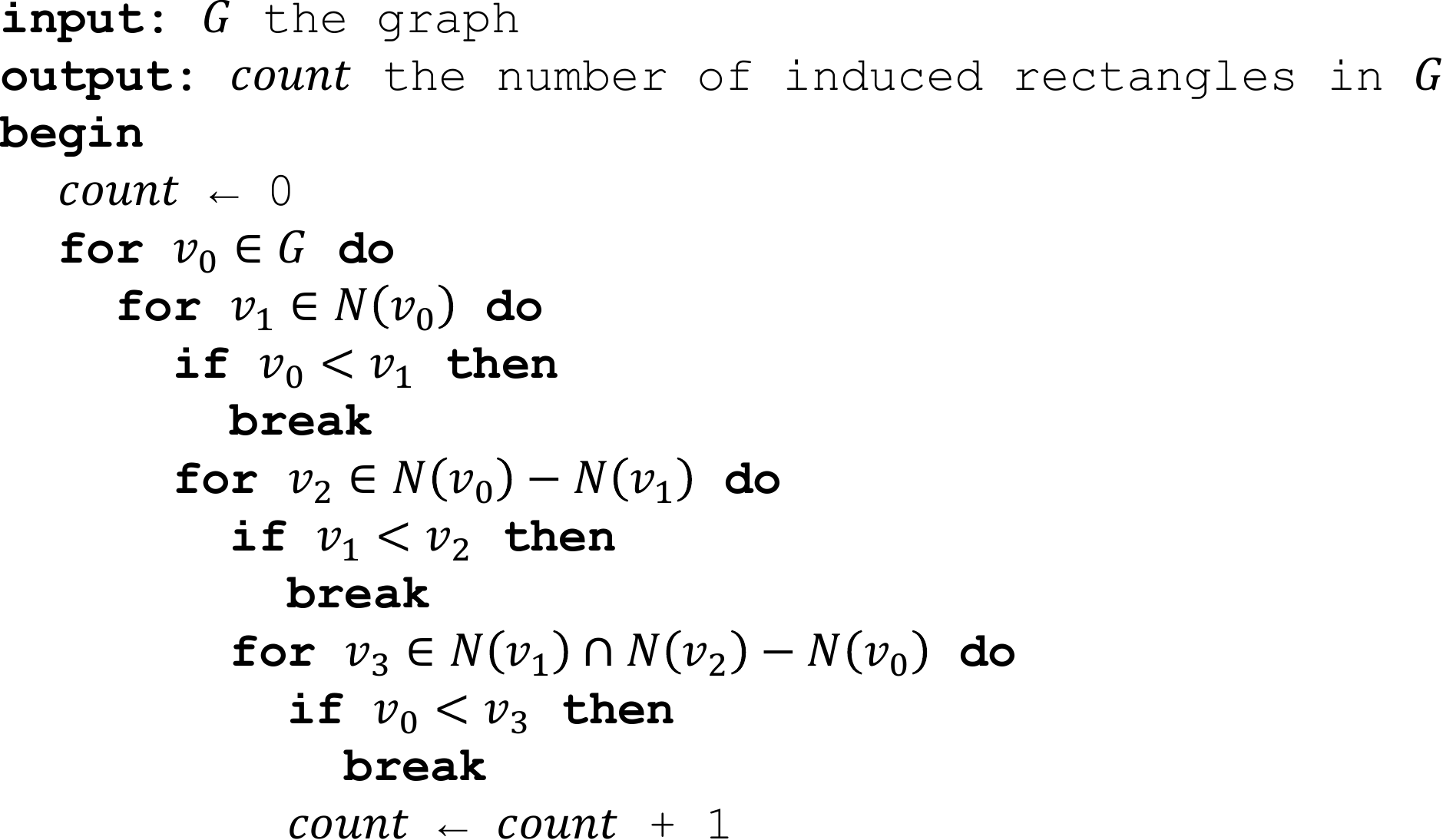}
        \caption{One Restriction per Vertex} 
    \end{subfigure}
    \caption{Rectangle counting with restrictions}
\label{fig:alg-rect} \end{figure*}

Figure~\ref{fig:alg-rect} (a) shows that when discovering one vertex the code may need to perform multiple checks, which may incur non-trivial overhead especially for large patterns.
We want to minimize the number of checks we need to perform to increase performance.
For example, within 3(a), the $v_0< v_2$ check is made redundant by the $v_0<v1$ check and the $v_1<v_2$ checks. 
These correspond to the $(A,C)$,$(A,B)$ and $(B,C)$ relations generated by the schedule, respectively.
The following theorem generalizes this process, and shows that we only need to enforce at most one restriction.
Hence, it is sufficient to perform at most one check when discovering a new vertex, and we only need to generate the relation corresponding to that check.

\begin{theorem}
\label{theorem:lazy_restrict} 
Given the set of binary relations generated by Algorithm~\ref{alg:multiplicity_reduce}, for each $k$ $(0 \leq k < |S|)$ where $S$ is the input schedule, we only keep $(S[z], S[k])$ where $z$ is maximized for $k$, we still properly enforce all the original binary relations. 
\end{theorem}

The proof is available in the appendix.
These relations impose restrictions on the IDs.
The key intuition arises when assuming that a vertex may be restricted above by two unrestricted vertices, which would require independent checks. 
The two restrictions imply that both vertices have an automorphism mapping them to the double-restricted vertex, otherwise the restrictions would not exist.
This implies the two unrestricted vertices would have an automorphism mapping between them, which implies a restriction between them, breaking the assumption.
Therefore, we only need to consider at most one relation bounding above any given vertex.

%

We can utilize Theorem \ref{theorem:lazy_restrict} to modify Algorithm~\ref{alg:multiplicity_reduce} to store a map of chosen relations. Instead of adding $(v,x(v))$ as a relation, we set the value in the map for $x(v)$ to be $v$, representing that the only relation that needs to be checked for $x(v)$ is $(v,x(v))$. 
Figure\ref{fig:alg-rect} (b) shows the generated code for rectangle counting after applying this optimization.







%



\section{Fast Schedule Generation}
\label{sec:generation}
In this section, we first explain the reason schedules of the same pattern have dramatically different performance.
We describe the schedule generation algorithm used in AutoMine and show why it incurs tremendous overhead, followed by the algorithm used in GraphZero's schedule explorer to efficiently prune the search space.
We finally present GraphZero's method of estimating performance of schedules bounded with restrictions.

\subsection{Performance Differences}
\begin{figure}
    \centering
    \includegraphics[width=0.25\textwidth]{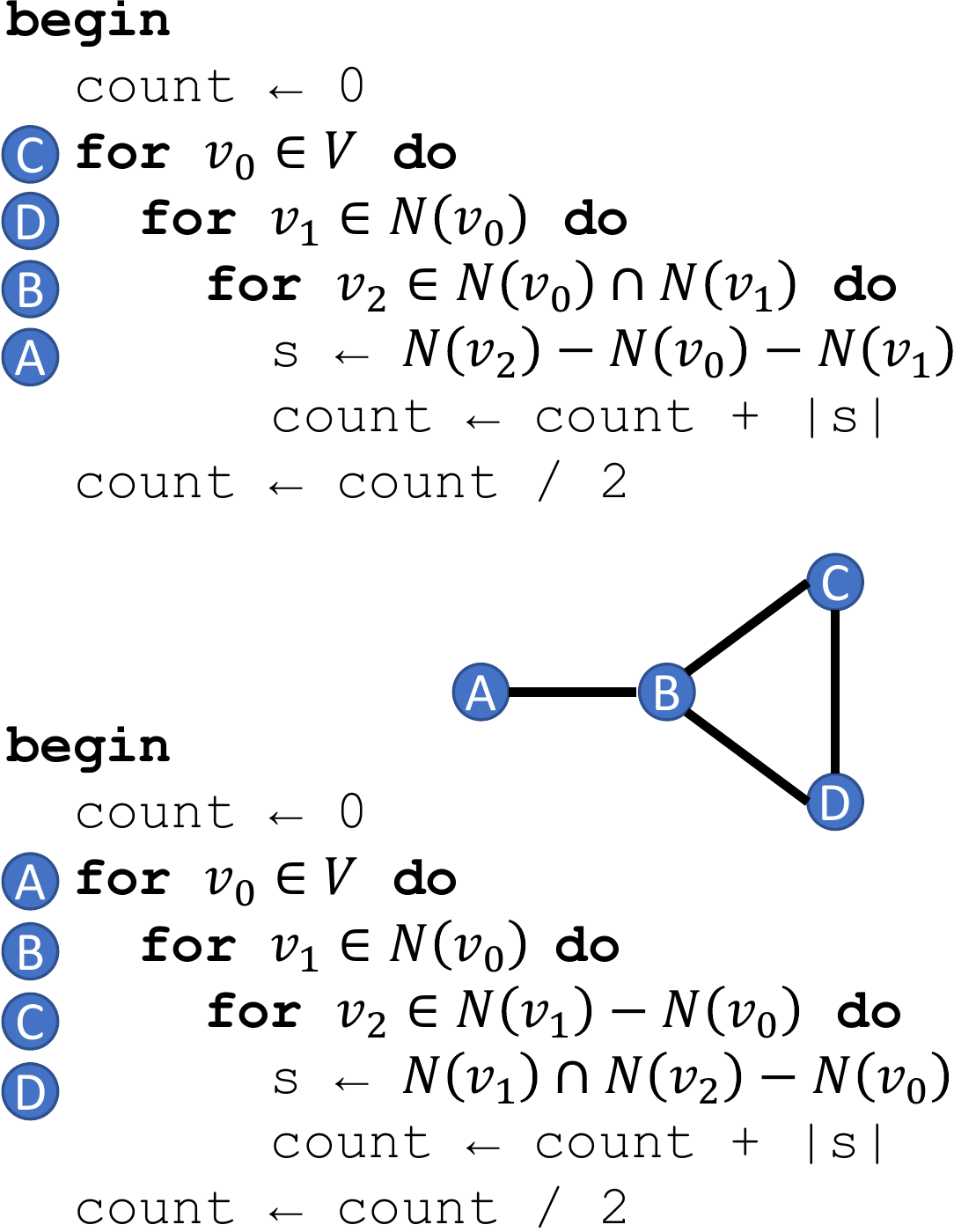}
    \caption{Tailed triangle counting with two different schedules}
    \label{fig:tailed}
\end{figure}

The performance of different schedules can vary significantly.
To illustrate this, we compare two schedules for tailed triangle.
As shown in Figure~\ref{fig:tailed}, the two schedules search for the vertices in $[A,B,C,D]$ and $[C,D,B,A]$ order according to the labels in the diagram. 
The key difference between these schedules is intuitive: the schedule $[C,D,B,A]$ executes its innermost loop once for each triangle embedding in the graph, while the schedule $[A,B,C,D]$ does it once for every wedge (i.e., a two-edge path). 
In some of the graphs we evaluate for this paper, wedges may appear over 500X more frequently than triangles. 
So if the amount of work done inside the innermost loop is comparable, the schedule $[C,D,B,A]$ should have much better performance.

\subsection{Brute-Force Schedule Generation}

To avoid missing high-performance schedules, AutoMine takes a conservative approach and tests if each permutation of the labeled vertices is a valid schedule as shown in Algorithm~\ref{alg:amsg}. Recall that a schedule is considered valid if each vertex, except for the first, is directly connected to at least one vertex that comes before it. This allows it to be described as a member of a composition of set differences and intersections of the neighbor sets of previous vertices. 
\begin{algorithm}
\SetKwInOut{Input}{input}
\SetKwInOut{Output}{output}
\SetKwProg{procedure}{Procedure}{}{}
\DontPrintSemicolon
\Input{$ $ $P:$ the pattern.}
\Output{$ $ $schedules:$ the list of all valid schedules}
\Begin{
\For{permutation $S$ of the vertices in $P$}{
\If{$S$ is a valid schedule}{
add $S$ to $schedules$}}}
\caption{Automine's schedule generation}
\label{alg:amsg}
\end{algorithm}

We can easily improve the schedule generation using Algorithm~\ref{alg:sched_gen_no_red}, which recursively searches only valid schedules. However, the algorithm is still inefficient for non-trivial patterns. To understand the complexity of traversing all valid schedules, we consider a path of length $n$. There are $n$ starting vertices to choose from and picking any vertex except the two end vertices leaves the remaining task as finding the valid schedules for the two sub-paths on the two sides. We omit the proof but just show that there exist a total of $2^{n-2}$ valid schedules. In another example, a clique of $n$ vertices, all $n!$ possible schedules are valid.

\begin{algorithm}
\SetKwInOut{Input}{input}
\SetKwInOut{Output}{output}
\SetKwProg{procedure}{Procedure}{}{}
\DontPrintSemicolon
\Input{$ $ $P:$ the pattern.}
\Output{$ $ $schedules:$ list of all valid schedules for $P$}
$schedules\gets []$\\
call recursive\_generate\_0([])\\
\procedure{recursive-generate\_0}{
    \Input{sched - a partially built schedule}
    \If{sched covers all vertices in the pattern}{add sched to schedules\\}
    \Else{\If{sched is empty}{
        $valid\_next \gets P$'s vertices}
    \Else{
    $valid\_next \gets$ all vertices in $P$ connected to sched but not in sched}
    \For{Vertex $v\in valid\_next$}{
       call recursive\_generate\_0(sched+[v])}}}
\caption{Schedule generation algorithm - smart brute force}
\label{alg:sched_gen_no_red}
\end{algorithm}

\subsection{Pruning the Valid Schedule Space}

The vast scheduling space described previously is expensive to search exhaustively, so we need an efficient method to prune the space without missing any potentially high-performance schedule.
A key observation we have is based on the clique example.
Although it has $n!$ schedules, all of them generate the same sequence of set intersection operations and hence generate the same code. We call such schedules equivalent schedules.
If two schedules are not equivalent, we call them distinct schedules.
In general, any two schedules are equivalent if and only if there is an automorphism that maps one to the other.
As an example, $(A,B,C,D)$ and $(A,B,D,C)$ are equivalent schedules for the tailed triangle in Figure~\ref{fig:tailed}. 
Recall that the multiplicity, $M$, of all valid schedules is the same because it is an inherent property of the pattern. Therefore, if $K$ schedules are valid for a pattern, there exist $K/M$ distinct schedules. The goal of the schedule explorer is to only generate these distinct schedules.
Algorithm~\ref{alg:sched_gen} describes GraphZero's schedule search approach which efficiently prunes the search space without missing high-performance schedules. The optimizations appear in two places: 1) valid schedule generation and 2) equivalent schedule pruning.

Each recursive invocation of the procedure $recursive\_generate$ tries to include one more vertex into the input schedule until the schedule covers all vertices in the pattern. 
To avoid generating invalid schedules, after the first vertex has been selected only vertices adjacent to at least one vertex already in the schedule are considered for extending the schedule (lines 9-12,23). 
These are the vertices contained within $valid\_next$.

The insight to prune the schedule space is similar to the idea used in Algorithm~\ref{alg:multiplicity_reduce}. 
When extending the partial schedule to include a vertex from $valid\_next$, selecting different vertices may produce equivalent partial schedules due to automorphisms. 
Algorithm~\ref{alg:sched_gen} leverages the automorphisms to partition $valid\_next$ into disjoint sets, such that for any two vertices $x,y$ in a set, any schedule generated by considering $x$ next is equivalent to one considering $y$ next. 
This property exists if and only if there is an automorphism remaining that moves $x$ to $y$.
Hence, the algorithm only expands the partial schedule by including the first vertex in each set and marks the rest as $processed$ (line 22). A $processed$ vertex is never considered to extend the partial schedule, thus pruning the schedule space.
\begin{algorithm}
\SetKwInOut{Input}{input}
\SetKwInOut{Output}{output}
\SetKwProg{procedure}{Procedure}{}{}
\DontPrintSemicolon
\Input{$ $ $P:$ the pattern.}
\Output{$ $ $schedules:$ the list of all distinct valid schedules for $P$}
\Begin{
    $schedules\gets []$\\
    $Aut \gets$ all the automorphisms of $P$\\
    call recursive\_generate([],Aut.\{\}, schedules)}
\procedure{recursive-generate}{    
    \Input{sched - a partially built schedule}
    \Input{aut - automorphisms that move $v$ to $v$ for $v\in sched$}
    \Input{valid\_next - list of potential next vertices}
    \Input{schedules - list of distinct schedules}
    \If{sched covers all vertices in the pattern}{
    add $sched$ with partial ordering $relations$ to schedules\\
    End this procedure call}
    \If{sched is empty}{
    $iterate\_over\gets P$}
    \Else{$iterate\_over\gets valid\_next$}
    \For{$v\in iterate\_over$ }{
        \If{$v$ is marked as processed}{
            continue}
        \Else{
            $stabilized\_aut\gets []$\\
            \For{$x \in Aut$}{
                \If{$x(v)= v$}{
                    add $x$ to $stabilized\_aut$}
                \Else{
                mark $x(v)$ as processed\\
                }}
            call recursive-generate(sched+[v], $stabilized\_aut$, $valid\_next  \cup N(v) - \{v\}$,schedules)\\ }}}
\caption{Schedule generation algorithm }
\label{alg:sched_gen}
\end{algorithm}


\begin{theorem}
\label{theorem:schedule_reduced.tex}
For a given Pattern, Algorithm~\ref{alg:sched_gen} generates all distinct schedules, and generates no two equivalent schedules. 
\end{theorem}

We include the proof in the appendix but show its basic idea. 
We can consider this proof to have two aspects.
The first that no two generated schedules are equivalent is obvious from the first vertex at which they differ. If one of these were processed, it would have marked the other.
The second, proving that every distinct schedule is generated, is equivalent to proving that every valid schedule is equivalent to one Algorithm~\ref{alg:sched_gen} generates. This is understood by noting that every possible trace of Algorithm~\ref{alg:sched_gen_no_red} is equivalent to one of the paths we explore with Algorithm~\ref{alg:sched_gen} because it never marks (or prunes) a vertex that does not have an indistinguishable vertex it has included in a partial schedule.
\subsection{Performance Model}
\label{sec:perf_model}
Now that we have successfully generated all distinct schedules, we need a performance model to select a high-performance schedule. Recall that once a schedule is generated, it is naturally mapped to a nested loop structure with restrictions as explained in Section~\ref{sec:reduction-intuition}. Figure~\ref{fig:alg-rect} (b) shows the nested loop structure with one restriction per nested for loop for rectangle counting. The performance model needs to estimate for each nested for loop 1) the number of iterations and 2) the number of iterations in which the restriction is satisfied. For the third for loop in the example, the performance model should estimate the number elements in $N(v_0)-N(v_1)$ and the number of times $v_1<v_2$ holds. 

The absolute numbers depend on the graph because in general the larger the graph is the bigger those numbers are. We therefore build a probabilistic model, assuming $n$ vertices in the graph and a probability $p$ for an edge to exist. For a for loop, given that its number of iterations is determined by $k1$ set intersections and and $k2$ set differences, the number of iterations is estimated as $np^{k1+1}(1-p)^{k2}$ by assuming that edge edge is equally likely to occur in a set operation. 

Note that given the estimation above, we only need to model the probability of satisfying the restriction. However, it is more difficult because the probability to satisfy the restriction in one for loop depends on the restrictions in all the for loops above it. We therefore model the probability of satisfying all checks up to the restriction of the considered for loop. 
The exact value of these probabilities is detailed within the appendix. 
Finally, we sum up the cost of all for loops to estimate the performance of the schedule.

\section{Generalizing Orientation-Based Optimization}
\label{sec:orientation}

In this section, we first explain why the orientation optimization is important for graph mining and the problems of generalizing it for arbitrary patterns. We then present GraphZero's method to leverage the automatically generated restrictions for the generalization.

Orientation optimization enforces an order to discover vertices by only allowing 
higher-degree vertices being discovered after lower-degree vertices. It works well for triangle
counting because every instance of the triangle pattern can be represented by a DAG,
starting from the lowest-degree vertex and ending with the highest-degree vertex.
The optimization allows pruning all edges from a higher-degree vertex to a lower-degree vertex but still guarantees that every triangle instance can be identified.

While the optimization only reduces the number of edges to process by a factor of 2, it substantially reduces the neighbor set size of “hot” vertices, which appear more frequently in patterns. Consider a star topology graph of
$N+1$ vertices with the center ``hot'' vertex connected to each of the other vertices.
With orientation optimization, its neighbor set becomes empty. But in the original graph,
the neighbor set of size $N$ needs to be accessed $N$ times.

Generalizing the orientation optimization faces two problems. First, the direction of the
pruned edge may interfere with the discovery order determined by the schedule. For instance, a restriction of the schedule may require that
two mapped vertices $v_i$ and $v_j$ should satisfy $v_i.ID > v_j.ID$, but $v_i$'s degree
is larger than $v_j$'s degree. In this case, the pruned edge is necessary for the schedule
to discover $v_j$ from $v_i$. Second, the orientation optimization must prune edges to gain
benefit, but the same graph may also be used by other applications which may not work with 
the modified dataset.

\begin{figure} \centering
\includegraphics[width=.45\textwidth]{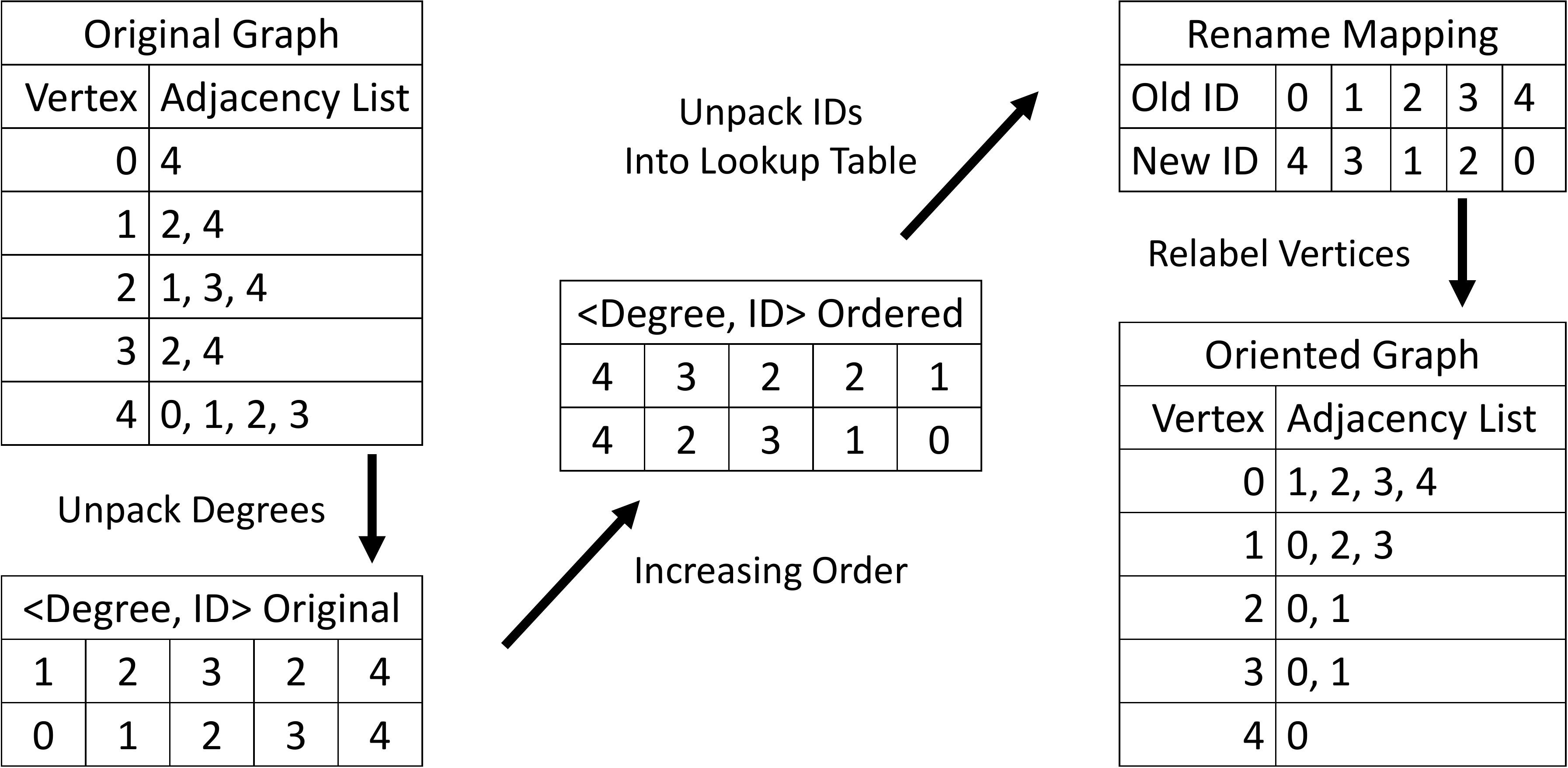}
\caption{Graph Reindexing Approach}
\label{alg:reordering} \end{figure}

\begin{figure} \centering
\includegraphics[width=.45\textwidth]{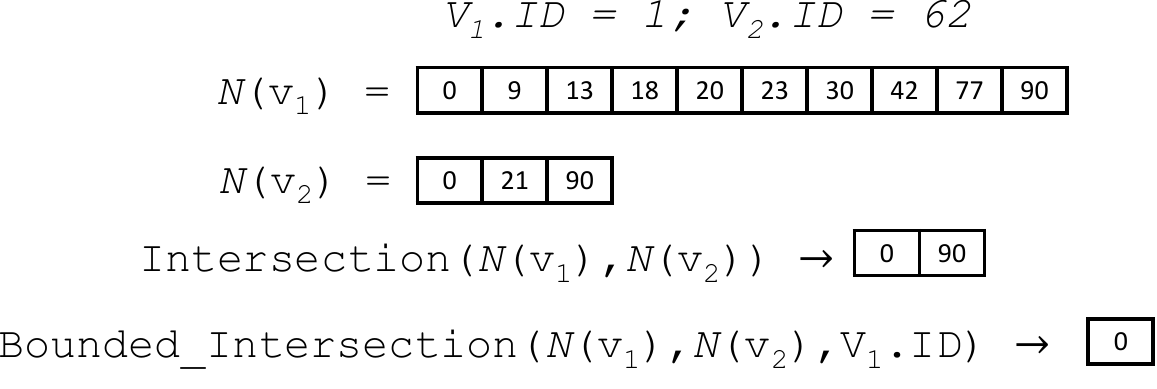}
\caption{Bounded Intersection Benefits}
\label{fig:intersection} \end{figure}

GraphZero generalizes the orientation optimization on arbitrary patterns without pruning
any edge by implementing two techniques.
The first technique reindexes the vertices such that the higher-degree vertices must have a smaller ID than that of the lower-degree vertices, as shown in Figure 5.
It uses the original IDs for tie-breaking when
vertices have the same degree. Recall that when GraphZero enforces a restriction $(v_i.ID >
v_j.ID)$, it also respects the discovery order from lower-degree vertices to higher-degree
vertices by only allowing $v_j$ to be discovered after $v_i$.
The second technique extends the set operations to leverage the restrictions for early exit (i.e., reducing the accessed elements in neighbor sets).
Given a restriction $(v_i.ID > v_j.ID)$, when discovering $v_j$ we are only interested in the neighbors in $v_i$'s neighbor set whose ID is less than $v_i.ID$. We can hence use $v_i.ID$ as a bound to perform the set operations. 
Because the neighbor sets are stored as sorted lists of integers, a linear scan finds the output vertices in sorted order, and can terminate whenever a bound condition is met.
The technique is particularly useful for large-degree vertices as shown in Figure~\ref{fig:intersection}. After the reindexing, $v_1$, a large-degree vertex, has a small ID as 1. If there exists a restriction involving $v_1$, the intersection can use $v_1.ID$ as the bound. In the example, the original intersection needs to traverse all elements in $N(v_1)$ while the bounded intersection only traverses three elements in total. 

\section{Evaluation}
\label{sec:eval}

In this section, we evaluate GraphZero and compare it with AutoMine, a state-of-the-art graph mining system that substantially outperforms other systems. 
The highlights of the results are as follows:
1) For 10 different workloads on real-world graphs, GraphZero is up to 40X faster than AutoMine running on the same system.
2) The compilation time reduction opens up the potential for a just-in-time compilation process with up to 197X speedup over the AutoMine compiler.
3) The generalized orientation optimization obtains the benefits of AutoMine's clique-specific optimization for arbitrary patterns yielding up to 88.6X speedup.

\subsection{Methodology}
\begin{table}
\small
	\begin{tabular}{c|c|c|c}
		\textbf{Graphs} & \#Vertices & \#Edges & Description \\
		\hline \hline
		CiteSeer~\cite{grami} & 3264 & 4536 & Publication citations \\
		Wiki-Vote~\cite{wikivote} & 7115 & 100762 & Wiki Editor Voting \\
		MiCo~\cite{grami} & 96638 & 1080156 & Co-authorship \\
		Patents~\cite{patents} & 3.8M & 16.5M & US Patents \\
		LiveJournal~\cite{livejournal} & 4.8M & 42.9M & Social network \\
		Orkut~\cite{orkut} & 3.1M & 117.2M & Social network \\
		Twitter~\cite{kwak2010twitter,BoVWFI,BRSLLP} & 41.7M & 1.2B & Social network \\
	\end{tabular}
	\caption{Graph Datasets}
	\label{tbl:datasets}
\end{table}

\begin{figure} \centering
\includegraphics[width=.4\textwidth]{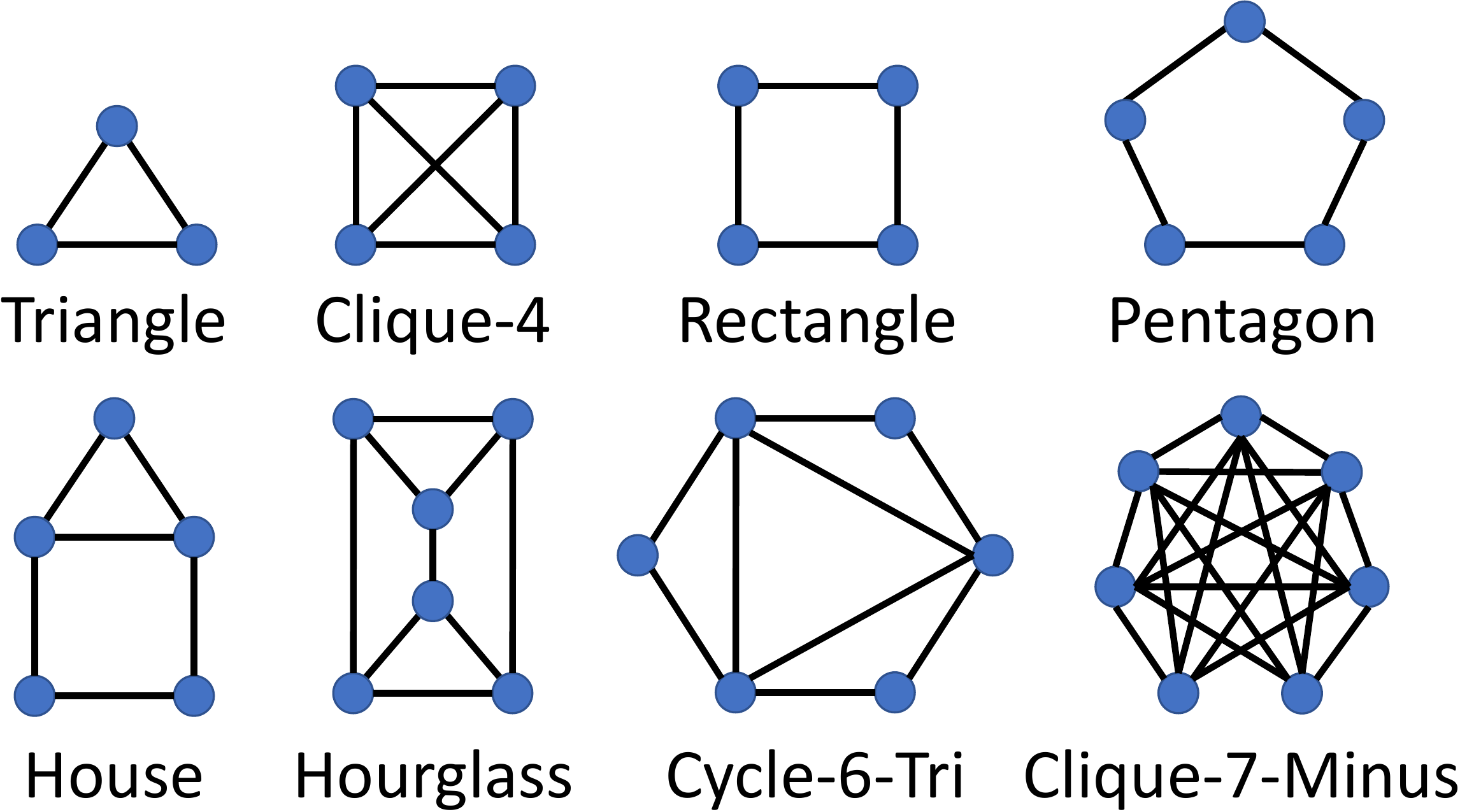}
\caption{Individual Pattern Workloads.}
\label{fig:patterns} \end{figure}

\paragraph{Experimental Setup}
Table \ref{tbl:datasets} shows the 7 real-world graphs used in the experiments.
AutoMine uses 5 of them to demonstrate that it outperforms prior mining systems including
Arabesque~\cite{arabesque} and RStream~\cite{rstream} by up to 4 orders of magnitude.
We hence also use these graphs to experiment with GraphZero.
We include Wiki-Vote to observe scalaibility to large patterns and Twitter to evaluate scalability to large graphs.
We run experiments on machines with 2 8-core Intel Xeon E5-2670 CPUs (hyperthreading disabled) and 64GB of memory.
Each machine runs Red Hat Enterprise Linux 6.9 with Linux kernel version 2.6 and gcc version 4.4.7, which we use with optimization level O3.
Figure \ref{fig:patterns} shows the 8 patterns we focus on for this evaluation, which range in size from 3 to 7 vertices.
We choose cliques of 3 (triangles) and 4 vertices, a near-clique of 7 vertices (missing
one edge), cycles of 4 (rectangle) and 5 (pentagon) vertices, then some interesting
patterns of 5 and 6 vertices.
We also perform motif counting, a popular application used to evaluate many other graph mining
systems~\cite{asap,rstream,automine}, on up to 5 vertices using aggregate schedules to efficiently count all motifs on a particular number of vertices.

\subsection{Performance Comparison}

\begin{figure*} \centering
\includegraphics[width=.9\textwidth]{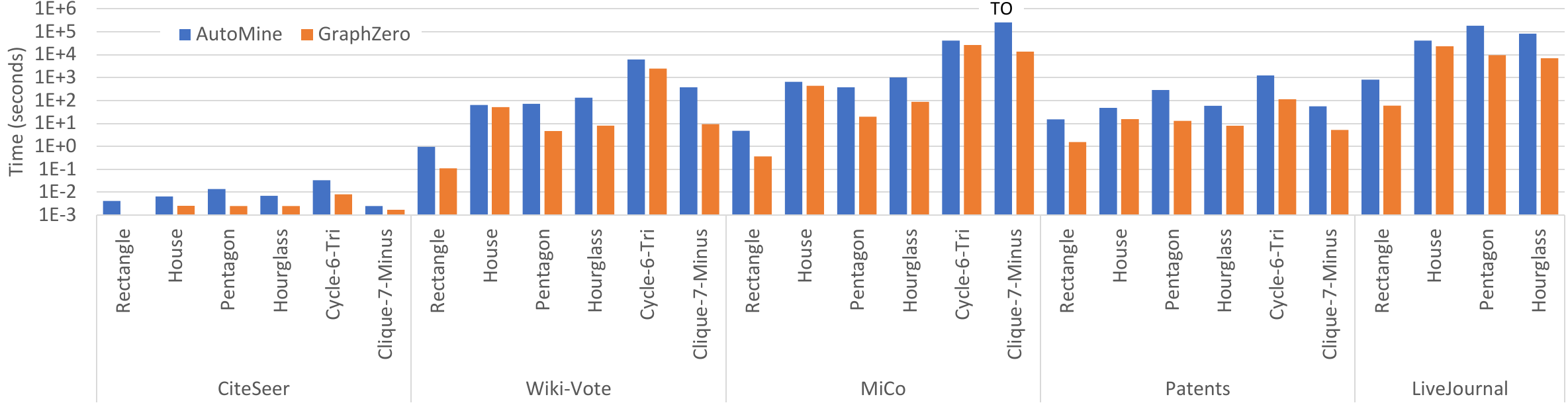}
\caption{Individual Pattern Performance ("TO" represents time out)}
\label{fig:individual} \end{figure*}

\paragraph{Individual Patterns}
We run 6 mining applications, corresponding to the non-clique target patterns in Figure \ref{fig:patterns}, on the 5 smaller graphs with both AutoMine and GraphZero. This directly showcases the improvements that multiplicity reduction offers. Notice that in Figure \ref{fig:individual}, GraphZero outperforms AutoMine by up to 40X for Clique-7-Minus and 22X for Pentagon, running on Wiki-Vote and Patents respectively. The best case scenario for those is expected to be 120X and 5X if the multiplicity reduction provided linear speedup, but GraphZero's generalized orientation optimization allows us to exceed that in the case of Pentagon.


\begin{table}
	\begin{tabular}{c|l|r|r}
		Graph & App. & AutoMine & GraphZero \\ \hline \hline
		\multirow{3}{*}{CiteSeer}
		& 3-MC & 1.6ms & 0.9ms \\
		& 4-MC & 11.9ms & 2.4ms \\
		& 5-MC & 537ms & 38ms \\ \hline
		\multirow{3}{*}{Wiki-Vote}
		& 3-MC & 34.5ms & 9.2ms \\
		& 4-MC & 11.5s & 1.7s \\
		& 5-MC & 5300s & 500s \\ \hline
		\multirow{3}{*}{MiCo}
		& 3-MC & 230ms & 60ms \\
		& 4-MC & 45.2s & 15.2s \\
		& 5-MC & 5.56h & 1.2h \\\hline
		\multirow{3}{*}{Patents}
		& 3-MC & 1.9s & 0.74s \\
		& 4-MC & 82.1s & 10.2s \\
		& 5-MC & 117m & 12.7m \\ \hline
		\multirow{2}{*}{LiveJournal}
		& 3-MC & 13.4s & 4.04s \\
		& 4-MC & 367m & 54m \\ \hline
		\multirow{2}{*}{Orkut}
		& 3-MC & 82.2s & 23.1s \\
		& 4-MC & 43.7h & 7.4h \\ \hline
		\multirow{1}{*}{Twitter}
		& 3-MC & 31.3h & 3.9h \\ \hline
	\end{tabular}
	\caption{Motif Counting Performance}
	\label{tbl:perf}
\end{table}

\paragraph{Motif Counting}
Motif Counting finds all connected patterns of a specified number of vertices, for our purpose between 3 and 5.
Both the complexity of each pattern and the number of patterns increase quickly with the number of vertices, with only 2 patterns on 3 vertices, but 21 on 5 vertices.
We run Motif Counting for 3, 4, and 5 vertices on all 7 graphs with a 72 hour timeout.
Table \ref{tbl:perf} shows all of the results that completed within the time limit. We observe that GraphZero outperforms AutoMine for all workloads with the smallest speedup of 1.75X and largest speedup of 14X. 
Notice that as the motif size increases, there is a corresponding sharp increase in the computational costs. We therefore only successfully finish the experiments on Motif-3 and Motif-4 with LiveJournal and Orkut and Motif-3 with Twitter. 
For the five smallest graphs, where the speedup is relatively small for Motif-3 (up to 3.8X), we see a trend of increasing speedup relative to AutoMine. On Motif-5, the speedup ranges from 4.7X to 14X.

\subsection{Schedule Generation}

\begin{figure} \centering
\includegraphics[width=.45\textwidth]{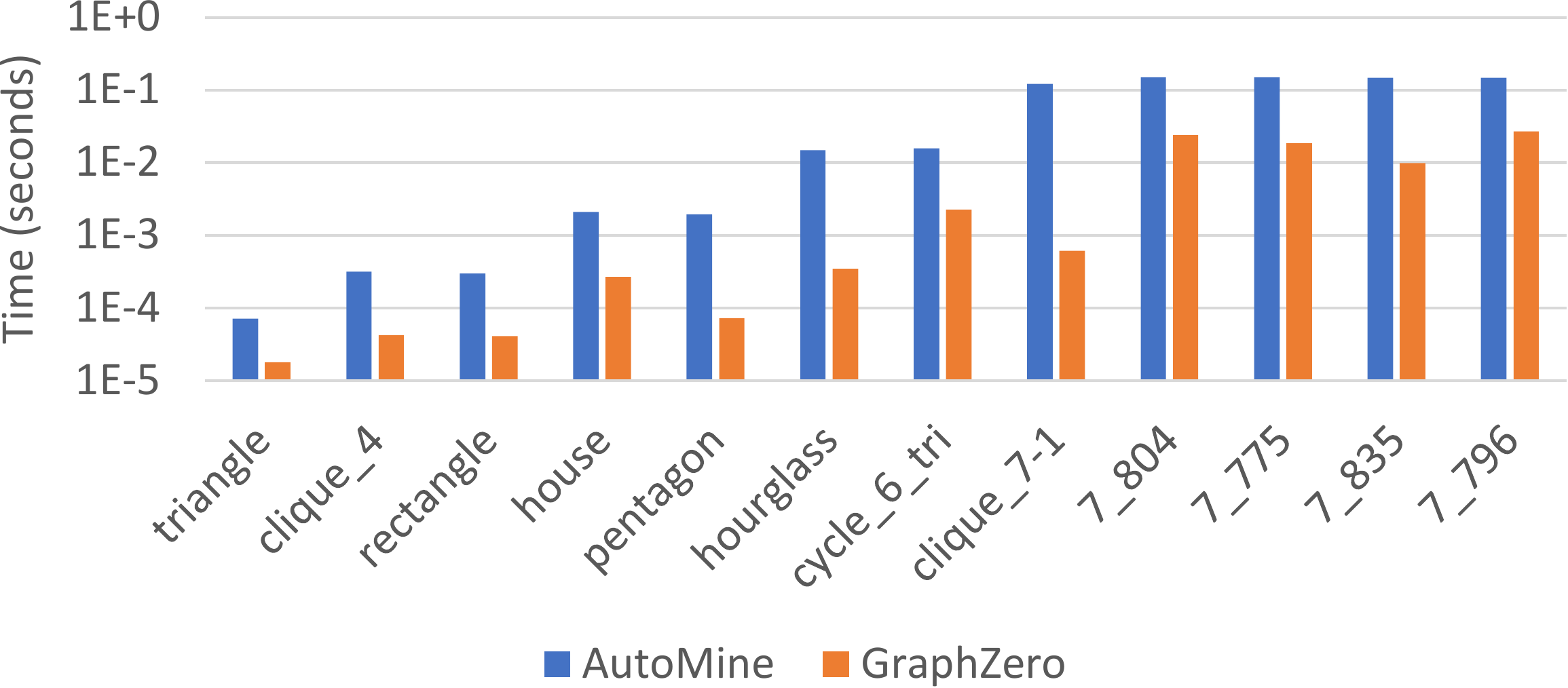}
\caption{Individual Pattern Compilation Time.}
\label{fig:compilation-individual} \end{figure}

\paragraph{Compilation Speed Comparison for Individual Patterns}
Considering patterns individually provides a direct comparison between the compilers in a pattern query scenario. Figure \ref{fig:compilation-individual} shows that where the AutoMine compiler takes up to 121ms for clique\_7-1, the GraphZero compiler takes just 0.6ms, a 197X speedup. Other 7-vertex patterns may take even longer. Of the 853 patterns of 7 vertices, the 4 most expensive (shown as 7\_804, 7\_775, 7\_835 and 7\_796) take over 149ms on their own in AutoMine. GraphZero completes all of these in under 30ms each, with an average speedup of over 8.7X and a maximum of 15.1X on the 7\_775 pattern (meaning index 775 in pattern discovery order). The results demonstrate that GraphZero's compilation technique has promise for use in a just-in-time compiler.

\paragraph{Compilation Speed Comparison for Motifs}
Combining the schedules for multiple patterns improves performance through data reuse, but can be expensive at compilation time. For Motif-7, each individual pattern is expensive to compile, and the aggregate compilation for all 853 patterns takes AutoMine over 2.5 minutes to complete. The speedup that GraphZero achieves on individual patterns naturally benefits the combined case, with a total compilation time of just 37 seconds, a 4X performance improvement. This trend will only continue with larger patterns, and with this performance, GraphZero has the potential to consider even larger patterns in the future.


\begin{figure} \centering
\includegraphics[width=.45\textwidth]{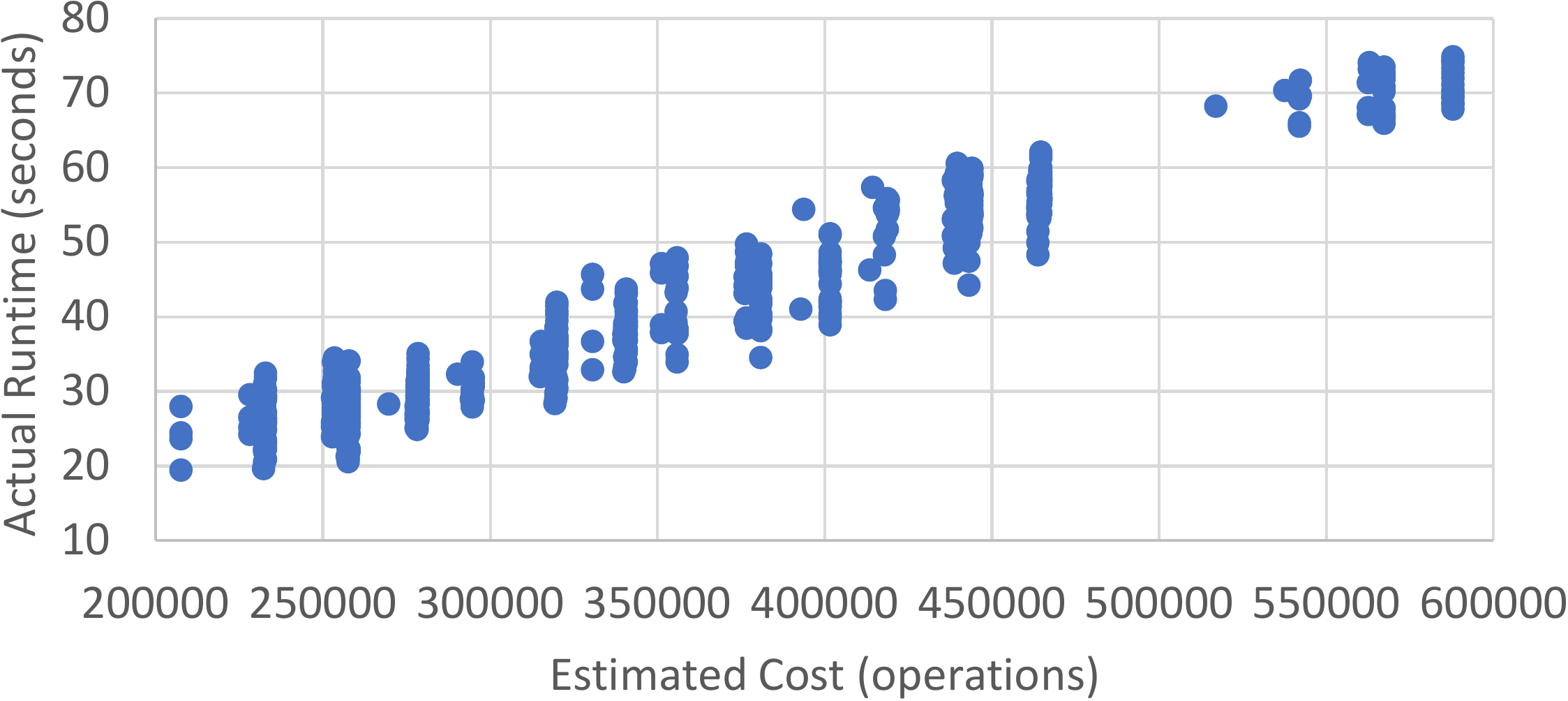}
\caption{Estimated versus Actual Performance on Patents}
\label{fig:correlation-patents} \end{figure}

\paragraph{Performance Modeling}
The performance model, as discussed in Section \ref{sec:generation}, estimates the number of operations each schedule needs to perform. It does not have to estimate real runtime, as its only purpose is to find the highest performance schedule. The estimates are computed in terms of number of operations performed in a uniform random graph with 1000 vertices and an average degree of 5. We show in Figures \ref{fig:correlation-patents} that the estimated cost and real performance are strongly correlated. The Coefficient of Determination describes the strength of statistical correlation, with 1 being a correlation that is perfectly defined by the data. We observe that for Patents this value comes out as 0.94, demonstrating the strong relative predictive capabilities of the performance model. The selected schedule according to the heuristic described in Section \ref{sec:perf_model} was the best schedule on Patents, though run-to-run variation can be up to 5\%. According to these results, we conclude that the performance model and heuristic successfully ensure that GraphZero selects a high-performance schedule.

\subsection{Orientation Optimization}

\begin{figure} \centering
\includegraphics[width=.45\textwidth]{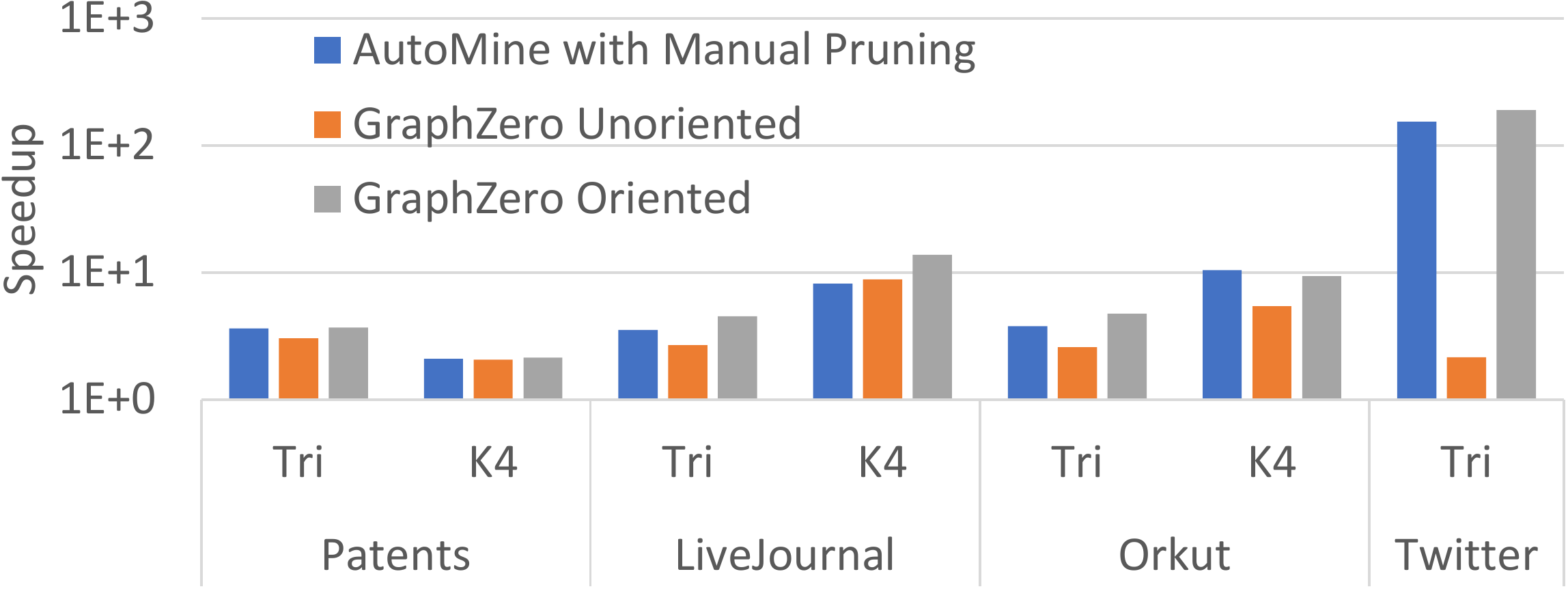}
\caption{Clique Performance Comparison.}
\label{fig:clique34} \end{figure}

\paragraph{Cliques}
The AutoMine work generalizes the orientation optimization to any-size cliques, but requires manually pruning all edges from higher-degree vertices to lower-degree vertices.
GraphZero's schedules automatically implement the optimization when running on the oriented graph.
Figure \ref{fig:clique34} uses the automatically generated code from AutoMine as the baseline and shows the speedups of AutoMine with manual pruning and GraphZero with both unoriented and oriented graphs on triangles and Clique-4.
We only show the four largest graphs in which triangle counting takes at least 100ms.
The manual pruning makes AutoMine much faster than the baseline, producing on average a 24.9X speedup for Triangles and 7.7X for Clique-4.
GraphZero without orientation achieves up to 94\% of its performance on MiCo Triangles, but falls behind in most other cases.
The oriented graph, however, allows GraphZero to achieve up to 1.4X and up to 2X speedup over AutoMine with Manual Pruning on Triangles and Clique-4, respectively, both with MiCo. We point out that GraphZero with orientation and AutoMine with manual pruning process the same amount of graph data (i.e., the edges from lower-degree vertices to higher-degree vertices). The extra performance benefit from GraphZero with orientation is from better load balance because of the degree-based sorting.

GraphZero's final performance beats default AutoMine by an average of 30.7X on Triangles and 9.4X on Clique-4.
It is especially interesting that GraphZero achieves a 192X speedup over the baseline for Twitter from using the orientation technique for Triangles.
A plausible reason is that because Twitter is the largest graph, its high-degree vertices contribute to a greater portion of the runtime than for other graphs. 

\begin{figure} \centering
\includegraphics[width=.45\textwidth]{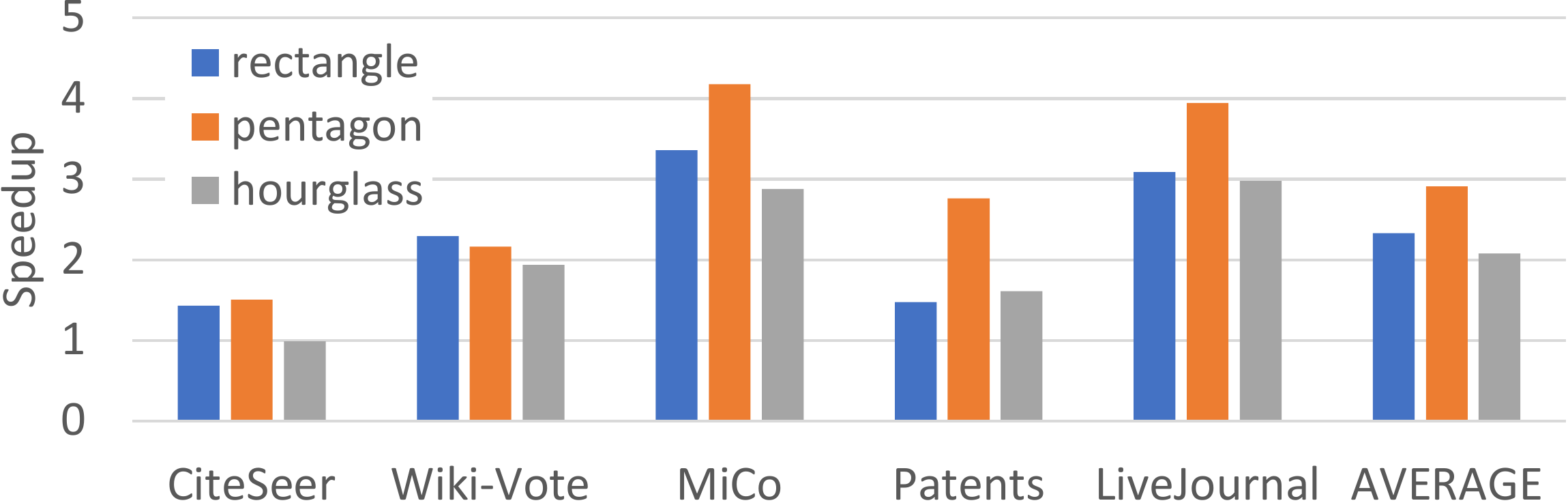}
\caption{Speedup from Orientation.}
\label{fig:orient-speedup} \end{figure}

\paragraph{General Patterns}
We evaluate the speedup directly attributed to the orientation in GraphZero for three patterns -- Rectangle, Pentagon, and Hourglass, on which AutoMine cannot apply the optimization.
Figure \ref{fig:orient-speedup} demonstrates the speedup of GraphZero with orientation over GraphZero without orientation.
The performance improves by up to 4.2X, with Pentagon seeing the most benefit at an average of 2.9X and Hourglass seeing the lowest benefit at an average of 2.1X.
These results are expected because the sparsity of Rectangle and Pentagon patterns makes them more likely to include high-degree vertices, whereas the dense subpatterns (triangles) in Hourglass filter out many of those high-degree vertices. But the benefit from orientation optimization is maximized on high-degree vertices. 


\begin{table}
\scriptsize
	\begin{tabular}{c|c|c|c|c|c|c}
	    CiteSeer & Wiki-Vote & MiCo & Patents & LiveJournal & Orkut & Twitter \\ \hline
	    3.3ms & 10.2ms & 32ms& 0.88s & 1.7s & 2.8s & 77.7s \\
	\end{tabular}
	\caption{Graph reindexing overhead}
	\label{tbl:orient}
\end{table}

\paragraph{Cost Benefit Analysis}
Using the oriented graph typically improves performance, but not necessarily for free.
Applying the orientation optimization in GraphZero incurs overhead for the reindexing process.
If the processing happens offline, there is no runtime overhead since the resultant graph is structurally equivalent to the original.
We also consider the possibility of reindexing the graph just-in-time, as the operations have low complexity compared to large pattern mining.
Table \ref{tbl:orient} reports the reindexing overhead for all the graphs used for evaluation. 
On CiteSeer, the smallest graph with only 4536 edges, even the 3.3ms time is large compared to the total mining time, so it is not worth applying the optimization. However, as the graph size increases, the benefit of orientation substantially outweighs the overhead. For the largest graph Twitter, reindexing takes 77.7 seconds, but saves over 4 hours of processing for Triangles. 
while the absolute cost to orient the graph indeed grows with the graph, the performance benefit scales up far faster.

\section{Related Work}
Many graph processing systems, including GraphLab~\cite{graphlab}, Graph\cite{graphchi}, Gemini~\cite{gemini}, Pregel~\cite{pregel}, GridGraph~\cite{gridgraph}, XStream~\cite{xstream}, and Ligra~\cite{ligra}, expose a
think-like-a-vertex or think-like-an-edge abstraction, which makes it easy to express
graph traversal algorithms such as breadth first search. The distributed systems focus on
optimizing communication~\cite{Shi:OSDI16}, locality~\cite{Gill:VLDB18}, and load balance~\cite{karypis1998multilevelk}, whereas the single-machine systems heavily
optimize I/O scheduling~\cite{wonderland}, minimize data loading~\cite{Vora:ATC16}, or trade off accuracy
for performance~\cite{Kusum:HPDC16}. However, none of these systems can be easily used to compose graph mining
applications, because of the gap between the low-level abstraction and the structural
patterns. 

Arabesque~\cite{arabesque} is the first distributed graph mining system that supports high-level interfaces
for user to easily specify and mine patterns. A number of graph mining systems are
then proposed with optimized memory consumption~\cite{gminer}, depth-first search~\cite{fractal}, out-of-core
processing~\cite{Kaleido,rstream}, or approximate mining support~\cite{asap,approxg}. As pointed out in
~\cite{automine}, these systems implement generic but inefficient mining algorithms and
incur unnecessary global synchronizations.

AutoMine is a unique graph mining system built upon a set-based representation. Given a
list of patterns, its compiler can automatically generate a nested loop structure of set
intersection and subtraction operations to identify all of them. It substantially
outperforms prior graph mining systems and is hence used as the baseline to 
evaluate GraphZero. EmptyHeaded~\cite{aberger2017emptyheaded} and GraphFlow~\cite{kankanamge2017graphflow} are
similar systems, but they focus on optimizing set intersection operations and cannot
handle missing edges in patterns. For example, they may consider the two-edge path in a
triangle as an instance of the wedge pattern, which is unacceptable in many applications.
In contrary, GraphZero, like AutoMine, supports arbitrary patterns and completely
eliminates redundancy in both schedule search and code execution.

Many compilers for graph analytics perform sophisticated optimizations once the graph
algorithm is clearly expressed using the provided domain-specific language.
GraphIt~\cite{graphit} enables user to describe the graph algorithm in the algorithm
language and how the algorithm should be optimized in the scheduling language. This 
separation allows the user to focus on algorithm design and offload the optimization tasks
to the compiler. Green-Marl~\cite{hong2012green}, SociaLite~\cite{seo2013socialite}, and
Abelian~\cite{gill2018abelian} can automatically parallelize and optimize graph algorithms but the
optimization space they can explore is significantly smaller compared with GraphIt. Pai and
Pingali~\cite{pai2016compiler} propose a set of compiler optimization techniques to efficiently map graph
algorithms to the GPU architecture. None of these compilers can generate efficient graph
mining algorithms, let alone remove redundancy, which is one of GraphZero's most important
contributions.

The orientation optimization was first proposed by Latapy~\cite{Latapy:TCS08} for triangle counting. Shun et al.~\cite{Shun:ICDE15} and Voegele et al.~\cite{tri} extend it for parallel triangle counting. Hu et al.~\cite{tricore} applies it to triangle counting on GPUs. AutoMine is the first work that generalizes the optimization for any-size clique patterns. To our knowledge, no prior work could apply the optimization to arbitrary patterns, which is made possible in GraphZero.





\section{Conclusion}

We proposed an optimizing compiler, GraphZero, that systematically addresses the limitations of AutoMine, a state-of-the-art graph mining system. GraphZero breaks symmetry in patterns, which causes serious performance and compilation overhead problems in AutoMine, through automatically generated and enforced restrictions. It leverages the generated restrictions and generalizes an important optimization for graph mining to arbitrary patterns based on a reindexing technique. The experiments showed that GraphZero substantially outperformed AutoMine for multiple patterns on real-world graphs with a much faster compilation speed.

\bibliographystyle{ACM-Reference-Format}
\bibliography{all}
\newpage
\appendix
\section{Appendix}
\begin{customthm}{1}\label{One}
For an arbitrary pattern $P$ and its schedule $S$, let $L$ be the partial ordering on the label set in $P$ generated by Algorithm~\ref{alg:multiplicity_reduce}.
Given an instance of $P$, denoted by $e$, in a graph $G$ there exists exactly one mapping $M$ from $P$ to $e$ if for each binary ordering relation $(S[i], S[j])$ in $L$, we add a restriction $M(S[i]).id > M(S[j]).id$. 

\end{customthm}
\begin{proof} This proof is split into two components. The first shows that at most one mapping exists that follows the ordering. The second shows that any instance that matches a mapping has at least one mapping that follows the restrictions.

To prove that at most one mapping exists, we perform a proof by contradiction.
Assume two distinct mappings, $M_1$ and $M_2$.
Then, $M_1$ and $M_2$ must differ in at least one vertex.
Let the first vertex in which they differ be $d$.
Then, there is an automorphism that doesn't change any of the vertices of $S$ which come before $d$, which maps $M_1$ to $M_2$. There is therefore an automorphism $A$ such that $M_1(A(x))= M_2(x)$.
Note that $M_2(d) \neq M_1(d)$, so $M_2(d) = M_1(A(d))< M_1(d)$ is a restriction that is enforced.
However, note that this symmetrically, we also have $M_1(d) < M_2(d)$.
These two form a contradiction, so there cannot be two distinct mappings that both obey the partial orderings. This concludes the proof that at most one is achieved.

To see that there is at least one mapping that satisfies the restrictions imposed by the binary relations, note that we can satisfy each of the restrictions in order by performing an automorphism to bring the maximum into the desired location at each step. There is an important property of these automorphisms that allow us to apply them without changing any relations, maintaining satisfcation of them, which is explained with the following lemma. 

\begin{customlemma}{1}\label{One}
Consider any automorphism $A$, which maps $S[i]$ to $S[i]$, for $i<k$.
$A$ does not change the set of binary relations $(S[a],S[b])$ where $a<k$, if we apply the automorphism to each binary relation $(X,Y)$ to become $(A(X),A(Y))$, over all automorphisms. As a consequence, applying such an automorphism does not break the relations.
\end{customlemma}
\label{lemma:Preserves}
\label{ax:lemma-Preserves}
\begin{proof}
First, for any $b<k$, the binary relations do not move, so they are automatically satisfied.
Now, consider $(S[a],S[b])$ as a binary relation, where $b\geq k$.
Then, there exists an automorphism that doesn't move $S[j]$ for $j<a$ that maps $S[a]$ to $S[b]$.
There is then an automorphism, which can be obtained by composition, which maps $S[a]$ to $A(S[b])$ which doesn't move $S[j]$ for $j<a$, which is the composition of those two automorphisms.
Hence $(S[a],A(S[b]))$ is a binary relation in the original set. 
Therefore, the set of binary relations after the automorphism is a subset of the binary relations before the automorphism. 
Then, symmetrically, since the automorphism is invertible with an automorphism, the before must be a subset of the after. 
Hence, the set of relations is the same before and after the automorphism is applied.
\end{proof}

That is, we can apply a sequence of $N$ automorphisms, and make a corresponding sequence of labelings, $M_0,M_1,M_2,\cdots M_{N}$.
Each member of these labelings will satisfy the following:
$M_j$ satisfies all relations of the form $(S[z], S[k])$, where $k<j$ is the 0-based index in the schedule, and $z$ is anything.
Note that we can allow $M_0$ to be $M$. We shall prove that we can construct this sequence iteratively.

We consider $M_k$ and construct $M_{k+1}$. Let $z$ be the index such that $M_k(S[z])$ is maximized, and where $(S[z],S[k])$ is a relation, or $z=k$.
Note that if $z=k$, we can just take $M_{k+1}=M_k$, as $M_k$ satisfies the restrictions, since $M_k(S[z])<M_k(S[k])$ for all $z$ where $(S[z],S[k])$ is a relation we have.
Otherwise, we do the following. There is an automorphism $A$ that doesn't change $M_k(S[x])$ for $x<k$, but takes $M_k(S[z])$ to $M_k(S[k])$.
This is required for the restriction to exist. 

We shall prove that letting $M_{k+1}=M_kA$, That is $M_{k+1}(x) = M_k(A(x))$ satisfies the restrictions we desired.
By the lemma, all relations $(S[a],S[b])$ where $b<k$ are still satisfied.

Now, to prove that $M_{k+1}$ satisfies the restrictions of the form $(S[x],S[k])$.
Note that $S[z] > S[d]$ for any $d$ where there is an automorphism mapping $S[k]$ to $S[d]$.
For any $x$ where such an automorphism exists, $(A(S[x]), S[k])$is also a restriction, since there is an automorphism mapping $ S[k]$ to  $A(S[x])$. 
Therefore, $M_{k+1}(S[x]) = M_k(A(S[x])) < M_k(S[z]) = M_{k+1}(S[k])$. Thus, $M_{k+1}$ satisfies the restrictions we desired. This completes the recursive step.

Therefore, we can construct $M_{N}$, which is a mapping that satisfies all the binary relations in the partial ordering.
Therefore, we have proved that it will find every distinct mapping at least once and at most once, so it finds each one exactly once, as desired.
\end{proof}

\begin{customthm}{2}\label{Two}
\label{ax:theorem-lazy-restrict}
Given the set of binary relations generated by Algorithm~\ref{alg:multiplicity_reduce}, for each $k$ $(0 \leq k < |S|)$ where $S$ is the input schedule, we only keep $(S[z], S[k])$ where $z$ is maximized for $k$, we still properly enforce all the original binary relations. 
\end{customthm}
\begin{proof}\label{proof:lazy_restrict} 
Note that for $(S[a],S[b])$ to be a binary relation in the partial ordering, $a<b$. 
This is obvious from the fact that no automorphism remaining at step $a$ can move $b$ if $b<a$, and $(S[a],S[a])$ cannot be a restriction.

We prove the theorem inductively. That is, if for $j< k$, we only enforce chosen relations of the form $(S[x],S[j])$ for $j$, that enforcing the chosen relation for $k$, if there is one, without considering the others, still enforces all relations for $j\leq k$.
The base case, where no restrictions need to be enforced, is trivial.
If there is no ordering $(a,k)$ then the recursive step is trivial. 
Note that if $(z,k)$ is required by the partial ordering, there is an automorphism mapping $z$ to $k$ that doesn't move $j$ for $j<z$. Let $(c,k)$ be the one the chosen relation for $k$. 
We know that for all $z\neq c$ such that $(z,k)$ is a restriction, $z<c$. 
Applying the chosen restriction obviously satisfies $(c,k)$ as a restriction. 
Note then that there is therefore an automorphism that takes $c$ to $k$, and thus one that takes $k$ to $c$, which doesn't change anything for $j<c$, and similarly doesn't change anything for $j<z$.
Then, since $(k,z)$ is a restriction, there is an automorphism that takes $z$ to $k$ that doesn't move $j<z$.
Hence, there is an automorphism taking $z$ to $c$ that doesn't move any $j<z$.
Thus, $(z,c)$ is also a relation, which is already enforced by the inductive hypothesis.
Since the restrictions are transitive, and $(z,c)$ and $(c,k)$ are both enforced, $(z,k)$ is enforced, as desired. 
Therefore, enforcing the chosen relations enforces all relations $(S[a],S[b])$ for $b\leq k$, completing the induction.
\end{proof}

\begin{customthm}{3}\label{Three}
\label{ax:theorem-schedule_reduced}
For a given Pattern, Algorithm~\ref{alg:sched_gen} generates all distinct schedules, and generates no two equivalent schedules. 
\end{customthm}

\begin{proof}
We first prove that it generates no two equivalent schedules. Say that it produces two equivalent schedules, $S_1$ and $S_2$.
Let the first vertex at which $S_1$ and $S_2$ differs be $V_1$ in $S_1$ and $V_2$ in $S_2$. 
There is necessarily an automorphism mapping $S_1$ to $S_2$.
When the algorithm produces the two schedules, at the point where $V_1$ and $V_2$ are processed, neither must be marked by the other. However, since there is an automorphism mapping $S_1$ to $S_2$, $V_1$ would mark $V_2$ and $V_2$ would mark $V_1$. This is a contradiction. Therefore it is impossible for two equivalent schedules to be generated by the algorithm.

Now, we must prove that every valid schedule is equivalent to one generated by the algorithm.
This becomes obvious if we can prove the following:
The automorphisms that move $v$ to $v$ for $v\in sched$ partition $valid\_next$ into disjoint sets, such that for any two vertices $x,y$ in a set, any schedule generated by considering $x$ next is equivalent to one considering $y$ next. 
When we consider the fact that these automorphisms form a permutation group, known as a stabilizer group, and let these partitions be the orbits of elements within the group. 
An automorphism mapping $x$ to $y$ maps any schedule with $x$ next to a schedule with $y$ next, and there is necessarily such an automorphism if they're within the same orbit.  
\end{proof}
\noindent {\bf Performance model}\\

To understand the probability evaluation, we model the probability that all restrictions only on vertices discovered in loops up to and including that layer are satisfied.
The probability of all the relations being satisfied is $\prod\frac{1}{z_i}$,
where $z_i$ is the number of vertices which are at most $v_i$ (including $v_i$), under restrictions imposed up to that layer.

The explanation of this quantity is as follows: 
If we assume $x_0,x_1,\cdots,x_k$ are uniformly randomly distributed variables in $(0,1)$, it is the probability that, for any binary relation we enforce in the schedule, $(S[i],S[j])$, where $0\leq i\leq j\leq k$
\[x_j<x_i\]
As established by Theorem~\ref{theorem:lazy_restrict}, each variable only needs one upper bound. We can then compute this probability with a series of nested integrals. 
At each point of integration, if we integrate in order from $x_k$ to $x_0$, the power of $x_i$ will be the number of $x_j$ where $j>i$ and $x_j<x_i$. This power is $z_i-1$, which integrates to $1/z_i x^{z_i}$. This naturally results in the quantity established above as the value of the entire nested integral.

For example, with the rectangle, of which 1/8th follow the restriction, there are 4 which are at most $v_0$, 2 which are at most $v_1$, and 1 each which are at most $v_2$ and $v_3$\\
\begin{align*}
    &\int_{0}^1\int_0^{x_0}\int_0^{x_1}\int_0^{x_0} 1 dx_3dx_2dx_1dx_0\\
&=\int_{0}^1\int_0^{x_0}\int_0^{x_1}x_0 dx_2dx_1dx_0\\
&=\int_{0}^1\int_0^{x_0}\int_0^{x_1}x_0 dx_2dx_1dx_0\\
&=\int_{0}^1\int_0^{x_0}x_0x_1 dx_1dx_0\\
&=\int_{0}^1x_0(x_0^2/2) dx_0\\
&= \frac{1}{4}*\frac{1}{2} 
\end{align*}
For the bottommost layer, the quantity expressed above is equal to the reciprocal of the multiplicity, as expected.
\end{document}